\documentclass[cameraready]{Interspeech}
\title{Lightweight Detection and Model Attribution of Synthetic Speech via Residual Statistical Fingerprints}

\author[affiliation={1}, orcid=0009-0000-9842-5856]{Matías}{Pizarro}
\author[affiliation={2}, orcid=0009-0002-6741-2432]{Mike}{Laszkiewicz}
\author[affiliation={3}, orcid=0000-0003-0678-3053]{Dorothea}{Kolossa}
\author[affiliation={1}, orcid=0000-0002-1916-7033]{Asja}{Fischer}
\address{
    $^1$ Faculty of Computer Science, Ruhr University Bochum, Germany \\
    $^2$ LKA NRW, Germany \\
    $^3$ Electronic Systems of Medical Engineering, Technische Universität Berlin, Germany
}

\email{Matias.PizarroBustamante@rub.de, dorothea.kolossa@tu-berlin.de, asja.fischer@rub.de}

\keywords{synthetic speech classification, model attribution, digital forensics, residual fingerprints}

\usepackage{comment}
\usepackage{tikz}
\usepackage{eso-pic}

\usepackage{cuted}

\begin{document}

\maketitle

% Noticia de aceptación fijada solo en la primera página
\begin{tikzpicture}[remember picture, overlay]
\node [
  anchor=south, % Alineado desde el borde inferior de la caja
  yshift=3cm  % Distancia desde el borde inferior del papel (ajústalo a tu gusto)
] at (current page.south) {
  \small\sffamily\itshape Paper accepted to Interspeech 2026 (Long Track). 10.21437/Interspeech.2026-1361
};
\end{tikzpicture}

% the abstract here must exactly match the abstract entered into the paper submission system
\begin{abstract}
    % 1000 characters. ASCII characters only. No citations.
    As speech generation technologies advance, so do risks of impersonation, misinformation, and spoofing.
    We present a lightweight, training-free method for synthetic speech detection and source model attribution.
    Our method builds on model-specific fingerprints
    that are computed as the average of the differences between audio signals and their filtered versions, referred to as residuals.
    Leveraging the Mahalanobis distance of the residual for a given audio signal to these model-specific fingerprints allows to identify the source model as well as to distinguish real from fake audio.
    A broad set of experiments across multiple synthesis systems and languages demonstrate a supreme performance of the proposed approach on four tasks: open-world single-model attribution, closed-world multi-model attribution, real vs.~synthetic speech classification, and out-of-domain detection.
\end{abstract}
\section{Introduction}
With the rapid advancement of synthetic data generation technologies, distinguishing between genuine and artificial speech signals has become increasingly challenging.
Public and open-source tools now enable even unskilled attackers to synthesize highly realistic voice samples that closely resemble human speech~\cite{ravanelli_2024_speechbrain}.
While these technologies offer clear benefits, such as improving availability for speech-impaired individuals~\cite{Metzger_2023_NatureJournal} and supporting multilingual communication~\cite{badlani2023_interspeech}, they also pose several security risks.
Spoofing attacks on automatic speaker verification systems, impersonation of trusted speakers, and misuse in voice-controlled interfaces are all well-documented threats. 
These threats can be exploited in various harmful ways, such as disinformation campaigns~\cite{Bontridder_Poullet_2021_article}, bypassing biometric authentication~\cite{Gupta_2024_EURASIP}, forging audio evidence in legal contexts~\cite{Delfino_2023_LawJournal}, or conducting identity fraud through voice cloning~\cite{klapsas_2022_interspeech}.
The emergence of synthetic audio as a vector for impersonation and manipulation poses a direct challenge to the security and trustworthiness of modern communication systems.
Current defenses largely focus on binary deepfake detection, i.e., deciding whether an audio signal is real or synthetic.
While this line of research is highly valuable and reflected in benchmark challenges such as Automatic Speaker Verification Spoofing (ASVspoof)~\cite{wang_2024_asvspoof5} and Audio Deepfake Detection (ADD)~\cite{Yi_2023_IJCAI}, it addresses only part of the problem.
Detection confirms manipulation, but it does not allow identification of the system responsible for the forgery.
This, however, is crucial across forensic, legal, and security domains.

Attribution matters for two reasons.
First, it provides actionable threat intelligence: linking an attack to a particular synthesis model reveals whether adversaries relied on commercial APIs, open-source implementations, or proprietary tools, thus informing risk assessment, replication potential, and mitigation strategies.
It also allows prosecutors to strengthen evidence chains by showing not just that an audio sample is fake, but also how it was generated.
Second, it underpins accountability and legal claims: rights holders can identify misuse of proprietary models.
The consequences of lacking attribution are evident in real-world incidents, e.g., in June 2025, adversaries used AI voice cloning to impersonate U.S. Secretary of State Marco Rubio, contacting foreign ministers via Signal and voicemail in an operation designed to extract sensitive information~\cite{bbc2025}.
While detection alone could flag the audio as fake, attribution would have offered critical insights about the tool employed, accelerating response and mitigation.
Compared to detection, model attribution (or generative source recognition) is still relatively underexplored, particularly in the context of audio.
Our work takes a step in this direction by showing how lightweight attribution can be both feasible and effective.

Existing speech synthesis attribution methods typically rely on multiclass classifiers trained in closed-world settings, limiting their ability to generalize to unseen models without retraining.
We instead propose a lightweight, training-free approach based on \emph{Residual Statistical Fingerprints} (RSFs), derived from standardized average residuals of the audio signal.
Attribution, classification, and out-of-domain tasks are all formulated as distance-based decisions using the Mahalanobis distance in the residual space.
The resulting framework is simple, adaptable, and well-suited to rapidly evolving speech generation technologies. 
Building on these insights, our key contributions are:

\noindent \textbullet\ \textbf{Residual speech fingerprints for attribution:}
Speech synthesis systems leave subtle, consistent residual artifacts, captured as model-specific RSFs.  

\noindent \textbullet\ \textbf{Unified distance-based framework:}
Model attribution and out-of-domain detection (OOD) tasks are formulated as Mahalanobis-distance decisions relative to the distribution of RSFs for each class.

\noindent \textbullet\ \textbf{Open-world single-model attribution efficiency:}
We determine whether a speech sample was generated by a given target model, rather than by an unseen synthesis system or a real speaker, without prior knowledge of all possible synthesis models.
Our method requires samples only from the target model during training and does not rely on data from all other synthesis systems.
%We introduce a single-model attribution technique in an open-world setting that determines whether a speech sample was generated by a given target model, rather than by an unseen synthesis system or a real speaker}. The method requires samples from only one target model \mb{and does not rely on collecting data from multiple known synthesis systems during training}.

\noindent \textbullet\ \textbf{Extensive evaluation:}
The approach is validated across diverse architectures, languages, and noise conditions, showing high performance and robustness to realistic audio distortions.

By addressing both attribution and classification of synthetic speech, this work advances lightweight, deployable defenses against malicious uses of generative models, contributing directly to applied cybersecurity and digital forensics.
\section{Related Work}
\label{sec:deepfake_detection}
We review prior work on synthetic speech detection and attribution, with emphasis on methods applicable to practical, automated identification of generative model outputs, and revisit the concept of model fingerprints originally proposed in the image domain for Generative Adversarial Networks (GANs).

\subsection{Synthetic Speech Detection}
These systems typically consist of an acoustic feature extractor and a classifier.
Common features include Linear-Frequency Cepstral Coefficients (LFCC)~\cite{Xinhui_2011_ASRU}, Mel-Frequency Cepstral Coefficients (MFCC)~\cite{Xinhui_2011_ASRU}, Constant Q~Cepstral Coefficients~\cite{Todisco_2017_JournalCSL}, and long-term variable-Q transforms~\cite{Li_2022_JournalDSP}.
These features are paired with classifiers such as Gaussian mixture models~\cite{ji_2017_interspeech} or neural models including X-vectors~\cite{Snyder_2018_SpokenLR}, Light Convolutional Neural Networks (LCNNs)~\cite{wu_2020_interspeech}, Residual Networks (ResNets)~\cite{Juntae_2023_ICASSP} and SE-ResNets~\cite{lai_2019_interspeech}, graph attention networks~\cite{tak_2021_interspeech}, differentiable architecture search~\cite{ge_2021_interspeech}, and Transformers~\cite{Xiaohui_2023_ICASSP}.
While highly effective for detection, these methods do not address model attribution.

\subsection{Synthesis System Attribution}
Model attribution aims to identify the specific generative synthesis system responsible for producing synthetic audio.
Deep learning-based synthesis systems outperform traditional parametric and concatenative methods, generating high-fidelity speech with improved prosody and speaker similarity.
Among the leading approaches are GAN-based~\cite{frank_2021_NeurIPS}, flow-based~\cite{Prenger_2019_ICASSP}, diffusion-based~\cite{huang_2022_IJCAI}, hybrid signal processing--neural models~\cite{wang_19_SSW}, and neural codec-based models~\cite{lu_2024_interspeech}.
These generative models—particularly GAN-based systems—have been studied in various classification and attribution works.
For example, Frank and Schönherr~\cite{frank_2021_NeurIPS} observed high-frequency artifacts in GAN-based speech synthesis.
Yan et al.~\cite{Xinrui_2022_DeepfakeWorkshop}, Li et al.~\cite{Fan_2024_ACM}, and Deng et al.~\cite{Junlong_2024_ICASSP} employed neural network-based classifiers, in some cases extended with contrastive learning, to attribute samples to known synthesis models.
Müller et al.~\cite{muller_2022_interspeech} introduced attacker-specific embeddings, while Stan et al.~\cite{stan_2025_interspeech} demonstrated kNN-based attribution using representations from pretrained self-supervised models.
However, these approaches rely on supervised learning with fixed label sets, making them inherently closed-world and requiring retraining when new models appear.

Recent work on open-world attribution and verification~\cite{falez_2025_interspeech, negroni_2025_interspeech} extends beyond binary detection, but largely relies on deep-learning pipelines requiring labeled data from multiple synthesis systems, with limited analysis of robustness, distribution shifts, and generalization to real speech.
Open-world evaluations such as ADD 2023~\cite{Yi_2023_IJCAI} still assume access to multiple known sources during training~\cite{Lu_2023_IJCAI, Qin_2023_IJCAI}, limiting applicability in single-model scenarios.
In contrast, our approach is training-free, does not require access to model parameters or architecture, and naturally supports open-world single-model attribution.
Related but complementary approaches include proactive audio perturbation techniques~\cite{Huang_2021_SLT, Zihao_2023_ACSAC}, which aim to prevent unauthorized synthesis before it occurs, and digital watermarking methods~\cite{cispa_2021_ICCV}, which embed ownership or provenance information directly into generated content.
While effective for prevention and model ownership verification, these approaches address different problem settings and are outside the scope of this work, which focuses on post-hoc, audio-only model attribution for forensic and investigative purposes.

\subsection{Revisiting GAN Fingerprints}
\label{subsec:related_work_fingerprint}
Model-specific fingerprints were first introduced by Marra et al.~\cite{marra2019gans} for image GANs.
They model a generated sample $x$ as the sum of semantic content $I(x)$ and a model-dependent fingerprint $F$, such that $x = I(x) + F$.
By suppressing semantic content through spatial filtering $f$, residuals $R = x - f(x)$ approximate the underlying fingerprint.
Model attribution is then performed by correlating the residuals of a test sample with an estimated fingerprint obtained by averaging residuals across samples.
Subsequent work has extended this concept primarily through supervised learning, often targeting specific model families such as GAN-based synthesizers~\cite{Ning_2019_ICCV, chuyuan-2024-chineseconf}, or by relying on paired real--synthetic data~\cite{Ke_2025_article}, which limits their applicability in open-world and single-model settings.
\section{Problem Settings and Methodology}
\label{sec:problem_setup}
We consider speech synthesis systems including text-to-speech (TTS)~\cite{wang_2017_interspeech}, voice conversion (VC)~\cite{kaneko_2017_interspeech}, and neural codec (NC)-based generators~\cite{lu_2024_interspeech}. 
These systems produce synthetic speech that may be used for benign or malicious purposes. 
The defender observes only audio signals and seeks to detect, attribute, or reject synthetic samples without access to the internal structure of the generative models.

\noindent \textbf{Attacker model:} We assume a non-adaptive adversary with access to one or more speech synthesis systems, either public or proprietary.
The attacker can generate audio that may undergo common channel distortions (e.g., noise, compression, reverberation).
The adversary does not know the defender’s detection or attribution strategy.

\noindent \textbf{Defender model:} The defender aims to detect synthetic speech or attribute it to its generating model.
Their capabilities vary across scenarios, from having access to samples from a single synthesis model to multiple generators.
Attribution is performed solely from the audio samples, without access to the internal workings of the generative models.
\subsection{Problem Settings}
\begin{figure}[t]
    \centering
    % Single-Model Attribution
    \begin{minipage}[t]{\columnwidth}
        \centering
        \includegraphics[width=\columnwidth]{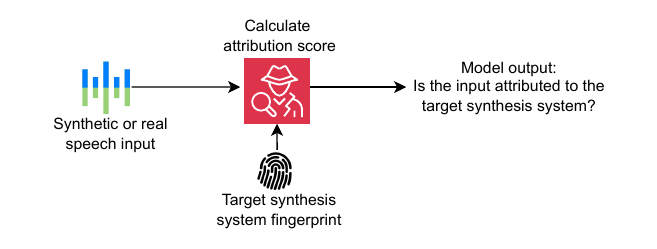}
        \footnotesize
        \parbox{\linewidth}{ \textbf{(a) Open-world single-model attribution:}
        Detects whether an audio sample has been generated by a specific target speech synthesis system.
        }
    \end{minipage}
    % Multi-Model Attribution
    \begin{minipage}[t]{\columnwidth}
        \centering
        \includegraphics[width=\columnwidth]{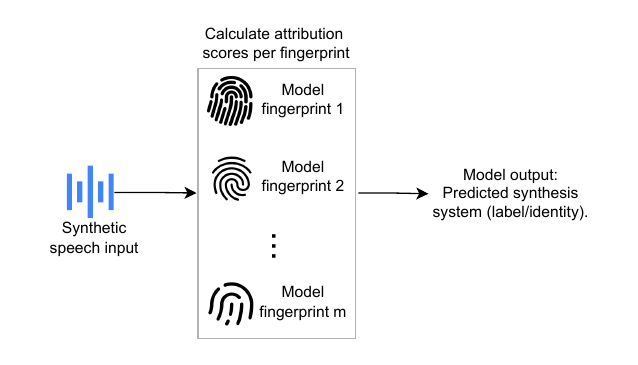}
        \footnotesize
        \parbox{\linewidth}{\textbf{(b) Closed-world multi-model attribution:} Classifies the synthetic audio as originating from one of several known models.}
    \end{minipage}
    \caption{Residual statistical fingerprinting system for model attribution: (a) open-world single-model attribution, and (b) closed-world multi-model attribution.}
    \label{fig:workflow_model_attribution}
\end{figure}

We define the attribution problem along two axes:
\begin{itemize}
    \item Open-world vs.~closed-world: Is the synthesis model used by the attacker assumed to be included in the construction phase of the attribution method, or can it be a previously unseen (unknown) model?
    \item Single-model vs.~multi-model: Data of how many synthesis models is available to the defender for constructing the attribution technique?   
\end{itemize}
These axes define four related tasks considered in this work:
\begin{enumerate}[label=(\alph*)]
    \item \textbf{Open-world single-model attribution (Fig.~\ref{fig:workflow_model_attribution}a):} Determine whether a given audio sample—potentially real or generated by an unknown system—was synthesized by a specific known model.
    \item \textbf{Closed-world multi-model attribution (Fig.~\ref{fig:workflow_model_attribution}b):} Identifying which model generated a synthetic sample when all candidate models are known.
    \item \textbf{Real vs.~synthetic classification:} Distinguishing bona fide (genuine) speech from synthetic audio.
    \item \textbf{Out-of-domain (OOD) detection:} Identify samples that do not conform to a set of known synthesis models.
\end{enumerate}
\subsection{Methodology}
\label{sec:main}
Fingerprinting approaches typically assume additive generative artifacts detectable via correlation~\cite{marra2019gans}, as is common in image data. 
In audio, however, artifacts are temporally distributed, spectrally correlated, and entangled with semantic content.
We introduce a model-agnostic
\emph{Residual Statistical Fingerprint} (RSF) framework that derives fingerprints directly from audio outputs without access to the generative model's architecture or parameters.
Our approach represents audio samples using fixed-length, time-averaged spectral representations, extracts residual embeddings through content-preserving filtering, and performs attribution using covariance-aware distance scoring.
The proposed framework addresses three core challenges specific to audio fingerprinting:
(I) variable-length signals,
(II) suppression of semantic speech content while preserving generative artifacts, and
(III) robust attribution via distribution-aware distance metrics.

\noindent \textbf{Challenge I: Mapping Variable-Length Audio Signals onto Fixed-Length Residuals:}
To enable consistent statistical comparison across audio samples of arbitrary length, we map each waveform to a fixed-length spectral representation.
Given an audio signal $x^{(i)}$, we compute its log-magnitude short-time Fourier transform (STFT) $\mathcal{S}_{x^{(i)}} \in \mathbb{R}^{F \times T^{(i)}}$, where $T^{(i)}$ denotes the number of time frames and $F$ the number of frequency bins.
Let $k \in \{1, \dots, F\}$ index the frequency bins, and define the time-averaged spectrum as: 
\begin{equation}
E_{x^{(i)}}[k] := \frac{1}{T^{(i)}} \sum_{t=1}^{T^{(i)}} \mathcal{S}_{x^{(i)}}(k,t) .
\end{equation}
This yields a fixed-dimensional vector $E_{x^{(i)}} \in \mathbb{R}^{F}$ that suppresses phonetic variability while retaining model-dependent spectral characteristics.

\noindent \textbf{Challenge II: Fingerprint Estimation via Content-Preserving Filtering:}
To isolate generative artifacts, we apply content-preserving filters for the residual computation.
The objective is to attenuate content-dominant structures while preserving synthesis-induced spectral traces.
We consider two complementary filtering strategies, applied independently:

\noindent (i) Neural Compression:
We use a pretrained EnCodec model operating at \SI{24}{\kilo\hertz} to compress and reconstruct each waveform~\cite{defossez_2023_TMLR}.\footnote{EnCodec pretrained model: \url{https://github.com/facebookresearch/encodec}.}
The learned quantization removes fine-grained details while preserving perceptual content, consistent with prior findings in the image domain showing that compression retains generative artifacts~\cite{Ning_2019_ICCV}.

\noindent (ii) Spectral Finite Impulse Response (FIR) filtering~\cite{oppenheim_2010_book}:
As an interpretable alternative, we apply FIR equiripple filters designed via the Parks–McClellan algorithm (MATLAB Filter Designer) to selectively attenuate frequency bands (low-pass, high-pass, band-pass, or band-stop), defined by the cutoff frequencies in Tab.~\ref{tab:spectral-summary}.
This design is motivated by prior observations that generative artifacts often appear in specific frequency regions, particularly at higher frequencies, as demonstrated by differences in spectrograms of real and synthetic speech~\cite{frank_2021_NeurIPS}.
Given a discrete-time input signal $x^{(i)}$, the filtered output $y^{(i)}$ is obtained via convolution with filter coefficients $\{h_k\}_{k=0}^{K-1}$:
$y^{(i)}[n] = \sum_{k=0}^{K-1} h_k\, x^{(i)}[n-k],$
where $n$ denotes the discrete-time sample index, $K$ the filter order, and zero-padding is applied at signal boundaries.

\begin{figure*}[t]
\centerline{\includegraphics[width=\textwidth]{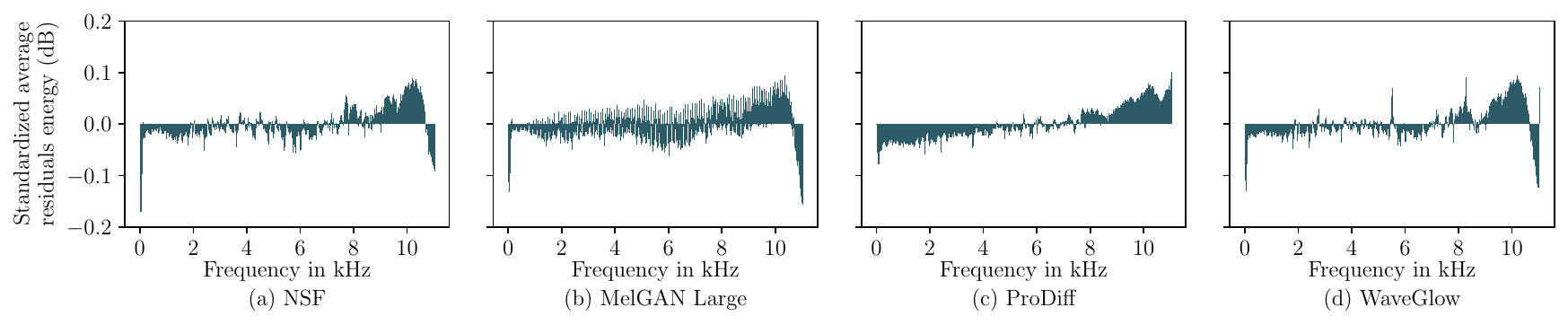}}
\caption{
Extracted RSFs \( \hat{\mathcal{F}} \) from four speech synthesis systems, computed as the mean residuals (Eq.~\ref{eq:residuals}) between the average spectral energy \( E_{x^{(i)}} \) of generated audio samples and their EnCodec-filtered counterparts \( E_{f(x^{(i)})} \), revealing model-specific spectral traits.
}
\label{fig:fingerprints}
\end{figure*}

Based on the filters, we estimate the fingerprints. 
It requires only samples from the target model and does not rely on real, paired, or out-of-domain data, enabling single-model and open-world attribution scenarios.
Given a filter $f(\cdot)$ and a set of $N$ audio samples $\{x^{(i)}\}_{i=1}^{N}$ generated by the same synthesis model, we define the residual embedding of each sample as
\begin{equation}
\mathcal{R}^{(i)} := E_{x^{(i)}} - E_{f(x^{(i)})} \in \mathbb{R}^{F}.
\label{eq:residuals}
\end{equation}
The RSF is then estimated as the empirical mean of residuals:
$\hat{\mathcal{F}} := \frac{1}{N} \sum_{i=1}^{N} \mathcal{R}^{(i)} .$
Fig.~\ref{fig:fingerprints} exemplarily illustrates the EnCodec-based fingerprints for four different speech synthesis systems.

\noindent \textbf{Challenge III: Distribution-Aware Scoring and Attribution:}
Prior fingerprinting approaches typically rely on correlation-based similarity measures~\cite{marra2019gans}, which ignore feature covariance and are sensitive to correlated spectral variations common in audio.
We instead treat residual embeddings as samples from a model-specific distribution and adopt the Mahalanobis distance for attribution.
Given a test sample residual $\mathcal{R}_{\mathrm{test}}$, its distance to a model fingerprint $\hat{\mathcal{F}}$ is computed as
\begin{equation}
d_{\mathrm{md}}(\mathcal{R}_{\mathrm{test}}, \hat{\mathcal{F}}) 
= \sqrt{(\mathcal{R}_{\mathrm{test}} - \hat{\mathcal{F}})^\top \Sigma^{-1} (\mathcal{R}_{\mathrm{test}} - \hat{\mathcal{F}})},
\end{equation}
where $\Sigma$ is the empirical covariance matrix estimated from the training residuals of the target model.
By normalizing deviations according to feature covariance, this distance measures the likelihood of a test sample under the target model’s residual distribution.
Such covariance-aware scoring is widely used in OOD detection~\cite{Podolskiy_2021_AAAI, Behrendt_2024_journal} and naturally supports open-world attribution without requiring access to non-target models during training.
Open-world single-model attribution uses a binary threshold on $d_{\mathrm{md}}$ to distinguish target from non-target samples, whereas closed-world multi-model attribution assigns a test sample to the closest model among a set of fingerprints $\{\hat{\mathcal{F}}_m\}_{m \in \mathcal{M}}$, where $\mathcal{M}$ denotes the set of known models:
\begin{equation}
\hat{m} = \arg\min_{m \in \mathcal{M}} d_{\mathrm{md}}(\mathcal{R}_{\mathrm{test}}, \hat{\mathcal{F}}_m).
\label{eq:naive_evasion_argmin}
\end{equation}
\section{Validation of RSF Design Choices}
\label{sec:validation}
This section validates key RSF design choices through controlled experiments on held-out data, assessing the suitability of distance metrics, spectral filters, and time–frequency representations rather than optimizing for benchmark-specific performance.

\subsection{Datasets}
To evaluate attribution across diverse synthesis techniques and deployment scenarios, we use multiple speech corpora covering different architectures, speakers, languages, and domains.

\noindent \textbf{Augmented LJSpeech (A-LJSpeech) Benchmark:}
We use LJSpeech~\cite{ljspeech_2017}, a single-speaker English corpus of $13{,}100$ utterances, as real speech. 
Synthetic samples are drawn from WaveFake~\cite{frank_2021_NeurIPS} and its extension~\cite{gasenzer_2024_TMLR}, which primarily include GAN-based vocoders.
To broaden coverage, we additionally generate speech using diffusion-based TTS models (FastDiff, ProDiff~\cite{huang_2022_ACM})\footnote{FastDiff, ProDiff pretrained models: \url{https://github.com/Rongjiehuang/ProDiff}.} and a hybrid neural source--filter (NSF) waveform model\footnote{NSF implementation: \url{https://github.com/nii-yamagishilab/project-NN-Pytorch-scripts}.} trained on LJSpeech.
For each real utterance, a synthetic counterpart is generated, yielding paired real--synthetic samples spanning GAN-, diffusion-, flow-, and hybrid-based synthesis. 
In total, A-LJSpeech includes speech from $10$ distinct synthesis models.

\noindent \textbf{JSUT Benchmark:}
The JSUT corpus~\cite{JSUT_2017}, a single-speaker Japanese dataset, is used to evaluate cross-lingual generalization.
The basic5000 subcorpus ($5{,}000$ utterances) provides real speech, while synthetic counterparts are generated using Parallel WaveGAN and Multiband MelGAN models from the WaveFake dataset.

\noindent \textbf{ASVspoof 2019 Logical Access (ASVspoof~LA) Benchmark:}
Derived from the VCTK corpus~\cite{veaux_2017_VCTK}, ASVspoof~LA~\cite{Wang_2020_JournalCSL} contains recordings from $107$ English speakers, including real speech (\emph{bonafide}) and synthetic speech from $19$ TTS/VC systems (122k utterances).
Training and development sets cover a subset of systems, with $11$ unseen systems in the evaluation set, providing a realistic multi-speaker benchmark.

\noindent \textbf{CodecFake Benchmark:}
It contains 1.06M utterances, including 132K real recordings in English (VCTK) and Chinese (AISHELL3)~\cite{shi_2021_interspeech}, and 926k synthetic samples generated by seven neural codec-based methods.
One generation method is reserved for testing cross-method generalization, enabling evaluation of attribution under codec-specific distortions.

\subsection{Distance Metric and Spectral Filter Selection}
\label{sec:spectral-config}
To validate our fingerprinting design, we conduct single-model attribution experiments on two held-out ASVspoof~LA systems (A16 and A19). 
For each system, $95$\% of samples were used to estimate RSFs, and $5$\% serve as a test set.
Attribution is evaluated by comparing residuals to model fingerprints using two scoring functions—correlation and the Mahalanobis distance—with residuals derived from multiple spectral filters (low-/high-pass, band-stop, band-pass) applied to STFT representations (\SI{8}{\milli\second} window, \SI{0.125}{\milli\second} hop).
Performance is measured using the Area Under the Receiver Operating Characteristic Curve (AUROC).
Results (Tab.~\ref{tab:spectral-summary}) show that the Mahalanobis distance consistently outperforms correlation, confirming that accounting for feature covariance is critical for reliable attribution.
Accordingly, the Mahalanobis distance is adopted for all subsequent experiments. 
Among spectral filters, the low-pass (\SI{1}{\kilo\hertz}) and band-pass (5–6~\si{\kilo\hertz}) configurations provide the most stable, discriminative residuals and are used throughout.%\footnote{See supplementary material for extended experimental details.}\label{fn:supp}
\begin{table}[!htbp]
\caption{
Comparison of distance metrics and spectral filtering strategies on ASVspoof~LA.
Values report average AUROC for RSFs A16 and A19 on held-out data.
For filters with multiple configurations, entries are shown in order of increasing cutoff or band (e.g., Low-pass: 1 / 3 / \SI{5}{\kilo\hertz}).$^{\dagger}$
}
\centering
\resizebox{\columnwidth}{!}{
\begin{tabular}{l|c|c}
\toprule
Filter type (configs) & Correlation & Mahalanobis distance \\
\midrule
Low-pass (cutoff: 1, 3, \SI{5}{\kilo\hertz}) &
0.91 / 0.92 / 0.92 &
\textbf{0.98} / 0.97 / 0.97 \\
High-pass (cutoff: 1, 3, \SI{5}{\kilo\hertz}) &
0.76 / 0.72 / 0.68 &
0.90 / 0.92 / 0.95 \\
Band-stop (bands: 4–7, 5–6 \si{\kilo\hertz}) &
0.67 / 0.59 &
0.94 / 0.91 \\
Band-pass (bands: 4–7, 5–6 \si{\kilo\hertz}) &
0.93 / 0.93 &
0.97 / \textbf{0.98} \\
\bottomrule
\multicolumn{3}{l}{$^{\dagger}$Bold denotes selected top filters; higher AUROC indicates better separation.}
\end{tabular}
}
\label{tab:spectral-summary}
\end{table}
\subsection{STFT Resolution Selection}
\label{sec:STFT-config}
We analyze the impact of time--frequency resolution on residual fingerprinting by comparing two STFT configurations: a fine-grained setting (\SI{8}{\milli\second} window, \SI{0.125}{\milli\second} hop) and a standard speech setting (\SI{25}{\milli\second} window, \SI{10}{\milli\second} hop). 
Tab.~\ref{tab:summary_comparison} reports average single-model attribution AUROC across benchmarks using the Mahalanobis distance with the selected low-pass (\SI{1}{\kilo\hertz}) and band-pass (5–6~\si{\kilo\hertz}) filters.
The finer resolution yields equal or slightly improved performance, particularly on ASVspoof~LA and A-LJSpeech, suggesting that finer temporal resolution better preserves synthesis-specific artifacts.
Accordingly, the \SI{8}{\milli\second} / \SI{0.125}{\milli\second} configuration is used in all subsequent experiments.%\footref{fn:supp}
\begin{table}[!htbp]
\centering
\caption{
Single-model attribution AUROC across benchmarks for two STFT settings using the Mahalanobis distance.
Values show low-pass \SI{1}{\kilo\hertz} / band-pass 5–6~\si{\kilo\hertz} scores (validation set).$^{\dagger}$
}
\resizebox{\columnwidth}{!}{
\begin{tabular}{lcccc}
\toprule
STFT (window / hop) & A-LJSpeech & JSUT & ASVspoof~LA & CodecFake  \\
\midrule
\SI{8}{\milli\second} / \SI{0.125}{\milli\second} & \textbf{1.00 / 1.00} & \textbf{1.00 / 1.00} & \textbf{0.99 / 0.99} & 0.98 / 0.98 \\ 
\SI{25}{\milli\second} / \SI{10}{\milli\second} & 1.00 / 0.99 & \textbf{1.00 / 1.00} & 0.95 / 0.97 & \textbf{0.98 / 0.98} \\ 
\bottomrule
\multicolumn{5}{l}{$^{\dagger}$ Bold marks best AUROC (strongest separation).}
\end{tabular}
}
\label{tab:summary_comparison}
\end{table}
\subsection{Content-Preserving Residuals: EnCodec vs. Spectral}
We compare two content-preserving residual constructions: residuals derived from EnCodec and those obtained via frequency-selective spectral filtering.
For both methods, residuals are computed from the same STFT representation (\SI{8}{\milli\second} window, \SI{0.125}{\milli\second} hop), ensuring a controlled comparison.
EnCodec residuals are evaluated using both correlation and the Mahalanobis distance, whereas spectral residuals use the Mahalanobis distance with the selected low-pass (\SI{1}{\kilo\hertz}) and band-pass (5--6~\si{\kilo\hertz}) filters.
As shown in Tab.~\ref{tab:encodec_vs_spectral_overall}, spectral filtering consistently achieves higher performance (AUROC over $0.98$) than EnCodec. 
The reduced EnCodec performance on CodecFake is likely due to samples generated by the EnCodec encoder itself, which diminishes residual discriminability.
These results indicate that frequency-selective filtering provides a more stable and interpretable residual representation for attribution.
\begin{table}[!htbp]
\centering
\caption{
Single-model attribution AUROC using content-preserving residuals.
EnCodec reports correlation / Mahalanobis scores; spectral filtering reports low-pass \SI{1}{\kilo\hertz} / band-pass 5--6~\si{\kilo\hertz} scores (Mahalanobis distance).$^{\dagger}$
}
\resizebox{\columnwidth}{!}{
\begin{tabular}{lcccc}
\toprule
Residual type & A-LJSpeech & JSUT & ASVspoof~LA & CodecFake \\
\midrule
EnCodec        & 0.99 / 0.95 & \textbf{1.00 / 1.00} & 0.95 / 0.89 & 0.85 / 0.86 \\
Spectral       & \textbf{1.00 / 1.00} & \textbf{1.00 / 1.00} & \textbf{0.99 / 0.99} & \textbf{0.98 / 0.98} \\
\bottomrule
\multicolumn{5}{l}{$^{\dagger}$ Bold marks best AUROC (strongest separation).}
\end{tabular}
}
\label{tab:encodec_vs_spectral_overall}
\end{table}
\section{Experimental Evaluation and Results}
\label{sec:evaluation}
Using the validated RSF configuration (Sec.~\ref{sec:validation}), we evaluate detection and attribution across multiple benchmarks, threat models, and robustness settings. 
Unless otherwise stated, benchmarks are split into $80$\% for training, $10$\% for validation, and $10$\% for testing.
To ensure fair comparison, all splits are class-balanced, with each category contributing an equal number of samples.
For the larger CodecFake benchmark, we use the size of the LJSpeech dataset as a reference and randomly sample that many examples per class from CodecFake.
Experiments are repeated five times and averaged.
All test sets contain unseen real and synthetic samples.
Code, data splits, and other resources are available at \url{https://github.com/matiuste/RSF}.

\subsection{Open-World Single-Model Attribution}
\begin{figure*}[t]
    \centering
    % Left panel (ProDiff)
    \begin{minipage}[t]{0.48\textwidth}
        \centering
        \includegraphics[width=\textwidth]{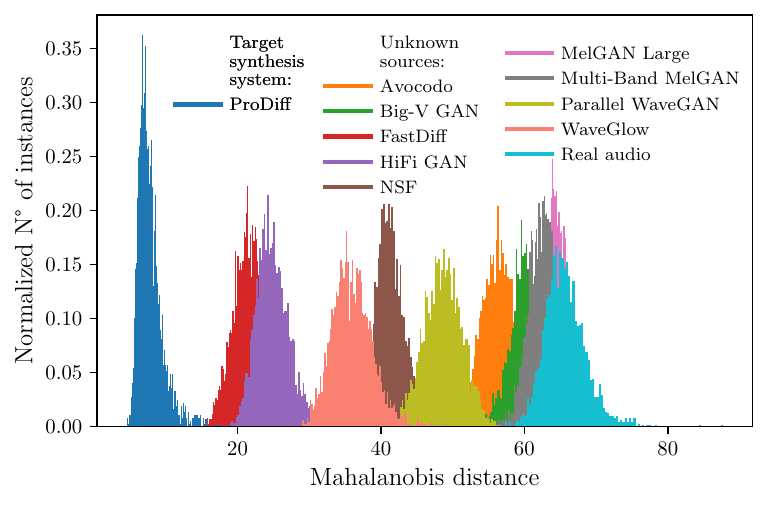}
        \\[0.3em]
        \footnotesize
        \parbox{\linewidth}{\textbf{(a) ProDiff model attribution:} Assessment by scoring unseen inputs from various sources with the Mahalanobis distance, including samples generated by the ProDiff model itself, using the A-LJSpeech benchmark.}
    \end{minipage}
    \hfill
    % Right panel (A02)
    \begin{minipage}[t]{0.48\textwidth}
        \centering
        \includegraphics[width=\textwidth]{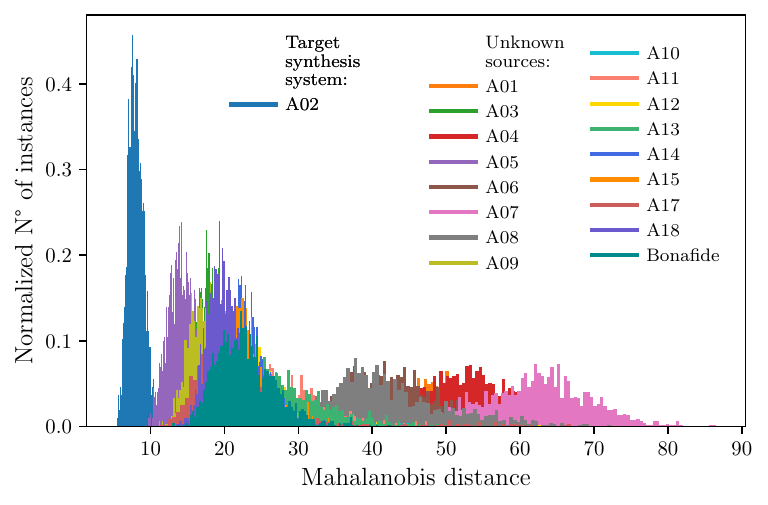}
        \\[0.3em]
        \footnotesize
        \parbox{\linewidth}{\textbf{(b) A02 model attribution:} Assessment by scoring unseen inputs from various sources with the Mahalanobis distance, including samples generated by the A02 model itself, using the ASVspoof~LA benchmark.}
    \end{minipage}
    \caption{Single-model attribution results for two target synthesis systems: (a) ProDiff and (b) A02. 
    Residual embeddings are extracted using a \SI{1}{\kilo\hertz} low-pass filter, with lower distances indicating that a sample is more likely to stem from the corresponding target system.}
    \label{fig:mahalanobis_combined}
\end{figure*}

\textbf{Experimental Setup:}
To determine whether an audio sample originates from a specific target synthesis system, we compare its residual vector $\mathcal{R}_{\mathrm{test}}$ with the known model RSF $\hat{\mathcal{F}}$ using the Mahalanobis distance.
Attribution is treated as a binary decision between target and non-target samples, and performance is evaluated via the AUROC. Fig.~\ref{fig:mahalanobis_combined} exemplarily shows attribution scores of unseen inputs—including real and synthetic samples—against a target RSF.

\noindent \textbf{A-LJSpeech \& JSUT:}
RSFs achieve near-perfect separation, with AUROC values consistently at $1.0$, demonstrating reliability across real and synthetic sources.

\noindent \textbf{ASVspoof~LA:}
Attribution remains strong across most systems (Tab.~\ref{tab:asvspoof_matrix}). 
Lower AUROC scores occur for acoustically similar systems: A01 and A04 are TTS systems sharing short-term acoustic patterns (cepstral, Mel-warped, and spectral-envelope features) used by their waveform generators ($0.86/0.87$); 
A06 and A04 rely on explicit waveform manipulations and MFCC features ($0.77/0.74$); 
and A02 and A03 use the same WORLD waveform generator and acoustic features ($0.79/0.82$). 
These reductions reflect intrinsic similarities rather than limitations of the RSF.
\begin{table}[!htbp]
\caption{Single-model attribution AUROC on ASVspoof~LA averaged over five runs.
Residual embeddings are compared against target RSFs using the Mahalanobis distance.
Rows denote source models and columns target models.
Results are shown for \SI{1}{\kilo\hertz} low-pass and 5–6~\si{\kilo\hertz} band-pass filtering.}
\centering
\resizebox{\columnwidth}{!}{
\begin{tabular}{lcccccc}
\toprule
Source\textbackslash{}Target & A01 & A02 & A03 & A04 & A05$^{*}$ & A06$^{*}$ \\
\midrule
A01 & - & 1.00 / 1.00 & 1.00 / 1.00 & 0.86 / 0.87 & 1.00 / 1.00 & 1.00 / 1.00 \\
A02 & 1.00 / 1.00 & - & 0.79 / 0.82 & 1.00 / 1.00 & 0.93 / 0.92 & 1.00 / 1.00 \\
A03 & 1.00 / 1.00 & 1.00 / 1.00 & - & 1.00 / 1.00 & 1.00 / 0.99 & 1.00 / 1.00 \\
A04 & 0.96 / 0.98 & 1.00 / 1.00 & 1.00 / 1.00 & - & 1.00 / 1.00 & 1.00 / 1.00 \\
A05$^{*}$ & 1.00 / 1.00 & 1.00 / 0.99 & 0.96 / 0.93 & 1.00 / 1.00 & - & 1.00 / 1.00 \\
A06$^{*}$ & 0.97 / 0.94 & 1.00 / 1.00 & 1.00 / 1.00 & 0.77 / 0.74 & 1.00 / 1.00 & - \\
A07 & 0.99 / 0.94 & 1.00 / 1.00 & 1.00 / 1.00 & 0.94 / 0.90 & 1.00 / 1.00 & 1.00 / 1.00 \\
A08 & 1.00 / 1.00 & 1.00 / 1.00 & 1.00 / 1.00 & 1.00 / 1.00 & 1.00 / 1.00 & 1.00 / 1.00 \\
A09 & 1.00 / 1.00 & 1.00 / 1.00 & 0.93 / 0.93 & 1.00 / 1.00 & 0.99 / 0.99 & 1.00 / 1.00 \\
A10 & 1.00 / 1.00 & 1.00 / 1.00 & 0.98 / 0.97 & 1.00 / 1.00 & 1.00 / 0.99 & 1.00 / 1.00 \\
A11 & 1.00 / 1.00 & 1.00 / 1.00 & 1.00 / 1.00 & 1.00 / 1.00 & 1.00 / 1.00 & 1.00 / 1.00 \\
A12 & 1.00 / 1.00 & 1.00 / 1.00 & 1.00 / 0.99 & 1.00 / 1.00 & 1.00 / 1.00 & 1.00 / 1.00 \\
A13$^{\ddagger}$ & 1.00 / 1.00 & 1.00 / 1.00 & 1.00 / 0.99 & 1.00 / 1.00 & 1.00 / 1.00 & 1.00 / 1.00 \\
A14$^{\ddagger}$ & 1.00 / 1.00 & 1.00 / 1.00 & 1.00 / 1.00 & 1.00 / 1.00 & 1.00 / 1.00 & 1.00 / 1.00 \\
A15$^{\ddagger}$ & 1.00 / 1.00 & 1.00 / 1.00 & 0.98 / 0.97 & 1.00 / 1.00 & 1.00 / 0.98 & 1.00 / 1.00 \\
A17$^{*}$ & 1.00 / 1.00 & 1.00 / 1.00 & 0.97 / 0.99 & 1.00 / 1.00 & 0.99 / 0.96 & 1.00 / 1.00 \\
A18$^{*}$ & 1.00 / 1.00 & 1.00 / 1.00 & 0.99 / 0.94 & 0.99 / 1.00 & 1.00 / 1.00 & 1.00 / 1.00 \\
Bonafide & 1.00 / 1.00 & 1.00 / 1.00 & 0.99 / 0.99 & 1.00 / 1.00 & 1.00 / 0.99 & 1.00 / 1.00 \\
\midrule
Avg. & 1.00 / 0.99 & 1.00 / 1.00 & 0.97 / 0.97 & 0.97 / 0.97 & 1.00 / 0.99 & 1.00 / 1.00 \\
\bottomrule
\multicolumn{7}{l}{Unmarked: TTS only; $^{*}$ VC systems; $^{\ddagger}$ TTS+VC; higher AUROC indicates better separation.}
\end{tabular}
}
\label{tab:asvspoof_matrix}
\end{table}

\noindent \textbf{CodecFake:} 
High AUROC is maintained across neural codec pairs (Tab.~\ref{tab:CodecFake_matrix}), with minor reductions for acoustically similar systems: 
C2 and C6 both use a Residual Vector Quantization (RVQ), producing similar patterns in certain frequency bands ($0.79/0.89$); 
C7, which spans a wide frequency range and uses enhanced RVQ with periodic activations, partially overlaps with C6 ($0.75/0.82$). 
These cases are due to shared architecture or quantization strategies.
\begin{table}[!htbp]
\caption{Single-model attribution AUROC on CodecFake averaged over five runs.
Residual embeddings are compared against target RSFs using the Mahalanobis distance.
Rows denote source models and columns target models.
Results are shown for \SI{1}{\kilo\hertz} low-pass and 5–6~\si{\kilo\hertz} band-pass filtering.}
\centering
\resizebox{\columnwidth}{!}{
\begin{tabular}{lcccccc}
\toprule
Source\textbackslash{}Target & C1 & C2 & C3 & C4 & C5 & C6 \\
\midrule
C1 & - & 1.00 / 1.00 & 1.00 / 1.00 & 1.00 / 1.00 & 1.00 / 1.00 & 1.00 / 1.00 \\
C2 & 1.00 / 0.99 & - & 1.00 / 1.00 & 0.99 / 0.99 & 0.99 / 1.00 & 0.79 / 0.89 \\
C3 & 1.00 / 1.00 & 1.00 / 1.00 & - & 1.00 / 1.00 & 1.00 / 1.00 & 1.00 / 1.00 \\
C4 & 1.00 / 1.00 & 1.00 / 1.00 & 1.00 / 1.00 & - & 1.00 / 1.00 & 0.99 / 0.99 \\
C5 & 1.00 / 1.00 & 0.96 / 0.95 & 1.00 / 1.00 & 0.99 / 0.99 & - & 0.91 / 0.94 \\
C6 & 1.00 / 0.99 & 0.87 / 0.87 & 1.00 / 1.00 & 0.99 / 0.99 & 0.99 / 0.99 & - \\
C7 & 1.00 / 1.00 & 0.93 / 0.95 & 1.00 / 1.00 & 0.99 / 0.99 & 1.00 / 1.00 & 0.75 / 0.82 \\
Real & 1.00 / 1.00 & 0.96 / 0.98 & 1.00 / 1.00 & 0.99 / 0.99 & 1.00 / 1.00 & 0.88 / 0.93 \\
\midrule
Avg. & 1.00 / 1.00 & 0.96 / 0.96 & 1.00 / 1.00 & 0.99 / 0.99 & 1.00 / 1.00 & 0.90 / 0.94 \\
\bottomrule
\multicolumn{7}{l}{$^{\dagger}$ Higher AUROC indicates better separation.}
\end{tabular}
}
\label{tab:CodecFake_matrix}
\end{table}

Overall, RSFs enable near-perfect attribution, reliably separating real from synthetic speech, and do so using only target-model samples for fingerprint estimation.
\subsection{Closed-World Multi-Model Attribution}
\label{sec:multi_model}
\textbf{Experimental Setup:}
The number of fixed candidate models per benchmark is $10$ for A-LJSpeech, $2$ for JSUT, $6$ for ASVspoof~LA, and $6$ for CodecFake.
We compare our approach against four strong baselines adapted for multi-model attribution: X-vector~\cite{Snyder_2018_ICASSP}, LCNN~\cite{lavrentyeva_2019_interspeech}, ResNet~\cite{Kaiming_2016_CVPR}, and SE-ResNet~\cite{Jie_2018_CVPR}.
All baseline models are trained for $10$ epochs using Adam (learning rate $0.001$, $\beta=(0.9,0.98)$) with a linear scheduler and $100$ warm-up iterations, using cross-entropy loss with softmax outputs.
The final prediction is the class with the highest probability.
Accuracy and F1-score are averaged over five runs.
These baselines are compared with our two RSF-based classifier, which operate directly on residual embeddings, which are extracted using a \SI{1}{\kilo\hertz} low-pass filter:

 \noindent \textbullet\ \textbf{RSF:} Assigns each sample to the synthesis system whose precomputed residual statistical fingerprint has the minimum Mahalanobis distance to the test residuals.
    No neural network training is required.
    
\noindent \textbullet\ \textbf{RSF CNN:} A compact CNN with squeeze-and-excitation modules, trained directly on residual features (STFT with \SI{25}{\milli\second} windows, \SI{10}{\milli\second} hop). 
    The network uses three convolutional blocks with increasing channel sizes ($32$, $64$, and $128$), each composed of $3\times3$ convolutions, batch normalization, and ReLU activations, with strided downsampling, followed by global average pooling and a linear classifier.
    It uses the same training protocol as baselines.

\noindent \textbf{Results:} Tab.~\ref{tab:closed-world} shows that RSF-based classifiers outperform or match the baselines across benchmarks. 
The RSF classifier achieves perfect accuracy on A-LJSpeech and JSUT, and near-perfect on ASVspoof~LA ($0.97$) and CodecFake ($0.99$). 
RSF CNN exceeds all other models, demonstrating that even a small CNN trained on residuals can match or surpass larger baselines.%\footref{fn:supp}
\begin{table}[!htbp]
\caption{
Closed-world multi-model attribution performance across benchmarks.
Cells report accuracy and F1-score, averaged over five runs with 10 training epochs each.$^{\dagger}$
}
\centering
\resizebox{\columnwidth}{!}{
\begin{tabular}{lcccc}
\toprule
Model & A-LJSpeech & JSUT & ASVspoof~LA & CodecFake \\
\midrule
X-vector        & 0.99 / 0.99 & 0.99 / 0.99 & \textbf{1.00 / 1.00} & \textbf{1.00 / 1.00} \\
LCNN            & 0.98 / 0.98 & 0.98 / 0.98 & \textbf{1.00 / 1.00} & 0.98 / 0.98 \\
ResNet          & 0.98 / 0.98 & 0.99 / 0.99 & \textbf{1.00 / 1.00} & \textbf{1.00 / 1.00} \\
SE-ResNet       & 0.98 / 0.98 & \textbf{1.00 / 1.00} & \textbf{1.00 / 1.00} & \textbf{1.00 / 1.00} \\
RSF (ours) & \textbf{1.00 / 1.00} & \textbf{1.00 / 1.00} & 0.97 / 0.97 & 0.99 / 0.99 \\
RSF CNN (ours)  & \textbf{1.00 / 1.00} & \textbf{1.00 / 1.00} & \textbf{1.00 / 1.00} & \textbf{1.00 / 1.00} \\
\bottomrule
\multicolumn{5}{l}{$^{\dagger}$ Bold marks best model performance.}
\end{tabular}
}
\label{tab:closed-world}
\end{table}
\subsection{Real vs. Synthetic Classification}
\label{real_vs_synthetic}
\textbf{Experimental Setup:}
All models from Sec.~\ref{sec:multi_model} are re-trained with binary cross-entropy for this task.
Logits are passed through a sigmoid function, and predictions are thresholded at $0.5$.
To maintain class balance, real speech samples are oversampled to match the number of synthetic ones.
Note that the test sets of ASVspoof~LA and CodecFake contain $11$ and $1$ unseen categories, respectively.
Thus, we also take the generalization capabilities into account.

\noindent \textbf{Results:}
Tab.~\ref{tab:fake-vs-real} reports accuracy and F1-score averaged over five independent runs.
Our classifier achieves perfect accuracy on A-LJSpeech and JSUT, and strong performance on ASVspoof~LA ($0.95$) and CodecFake ($0.94$), highlighting that RSF-based classification matches or surpasses the performance of significantly more complex models trained on the raw input.%\footref{fn:supp}
\begin{table}[!htbp]
\caption{
Real vs.~Synthetic classification performance across benchmarks.
Cells report accuracy and F1-score, averaged over five runs with 10 training epochs each.$^{\dagger}$
}
\centering
\resizebox{\columnwidth}{!}{%
\begin{tabular}{lcccc}
\toprule
Model & A-LJSpeech & JSUT & ASVspoof~LA & CodecFake \\
\midrule
X-vector        & 0.91 / 0.92 & \textbf{1.00} / \textbf{1.00} & 0.87 / 0.86 & 0.93 / \textbf{0.93} \\
LCNN            & 0.96 / 0.96 & 0.99 / 0.99 & 0.93 / 0.93 & 0.93 / \textbf{0.93} \\
ResNet          & 0.98 / 0.98 & \textbf{1.00} / \textbf{1.00} & \textbf{0.95} / 0.94 & \textbf{0.94} / \textbf{0.93} \\
SE-ResNet       & 0.98 / 0.98 & \textbf{1.00} / \textbf{1.00} & \textbf{0.95} / 0.94 & \textbf{0.94} / \textbf{0.93} \\
RSF CNN (ours)     & \textbf{1.00} / \textbf{1.00} & \textbf{1.00} / \textbf{1.00} & \textbf{0.95} / \textbf{0.95} & \textbf{0.94} / \textbf{0.93} \\
\bottomrule
\multicolumn{5}{l}{$^{\dagger}$ Bold marks best model performance.}
\end{tabular}
}
\label{tab:fake-vs-real}
\end{table}
\subsection{Out-of-Domain Detection}
\textbf{Experimental Setup:}
We use the ASVspoof~LA benchmark, treating A01--A06 as in-domain (ID) classes and randomly selecting six from the remaining systems as OOD candidates.
All classifiers are trained only on the six ID classes, while the unseen systems are reserved for OOD evaluation.
For OOD scoring, baselines use a confidence-based criterion: the OOD score is defined as one minus the maximum predicted probability over the ID classes, so lower confidence yields higher OOD scores.
We additionally incorporate as another baseline the recent TADA framework~\cite{stan_2025_interspeech}, which employs a pretrained wav2vec-based self-supervised model~\cite{Baevski_2020_NEURIPS} to compute audio features, and then adopt a $k$-NN model ($k=21$) based on Euclidean distance.
In our RSF approach, the OOD score is the minimal Mahalanobis distance between a test sample's residual embedding and the set of ID fingerprints, where larger distances indicate greater deviation from known ID systems.

\noindent \textbf{Results:}
Tab.~\ref{tab:ood_results} shows that the proposed RSF approach outperforms all baselines across both the AUROC and the area under the Precision-Recall Curve (AUPRC).
Specifically, RSF achieves $0.98$ on both metrics, surpassing TADA and substantially improving over the neural classifiers.
In contrast to the baselines, which require training a dedicated classifier on the ID classes, RSF relies solely on fingerprint estimation.
When a new synthesis system is introduced, no model retraining is needed; only the corresponding fingerprint must be computed and added to the ID set.
This makes the approach computationally efficient and scalable, aligning with Green AI principles by avoiding repeated training as the number of ID classes grows.
\begin{table}[!htbp]
\caption{
OOD performance on ASVspoof~LA.
Rows report baseline classifiers and our RSF method (residuals extracted with a \SI{1}{\kilo\hertz} low-pass filter).
Values show mean $\pm$ standard deviation over five runs for AUROC and AUPRC (higher is better), with best results in bold.
}
\centering
\footnotesize
\begin{tabular}{lcc}
\toprule
Method & AUROC & AUPRC \\
\midrule
X-Vector 
& $0.89 \pm 0.01$ 
& $0.89 \pm 0.01$ \\

LCNN 
& $0.79 \pm 0.10$ 
& $0.81 \pm 0.08$ \\

ResNet 
& $0.93 \pm 0.01$ 
& $0.94 \pm 0.01$ \\

SE-ResNet 
& $0.93 \pm 0.03$ 
& $0.94 \pm 0.02$ \\

TADA 
& $0.97 \pm 0.01$ 
& $0.96 \pm 0.01$ \\

RSF (Ours) 
& $\mathbf{0.98 \pm 0.01}$ 
& $\mathbf{0.98 \pm 0.01}$ \\
\bottomrule
\end{tabular}
\label{tab:ood_results}
\end{table}
\subsection{Efficiency and Amount of Data Needed}
We study the impact of training data size on RSF extraction by using subsets of $70$, $80$, $90$, $100$, $110$, and the full training set for each benchmark.
Single-model attribution is then evaluated on the full test set.
Tab.~\ref{tab:sample_efficiency} reports AUROC averaged across all source--target pairs. 
Attribution remains robust even with few samples: A-LJSpeech reaches near-perfect AUROC with $70$ samples and achieves $1.00$ from $80$ samples onward; JSUT improves from $0.96$ at $70$ samples to $1.00$ at $100$; ASVspoof~LA and CodecFake achieve stable AUROC with $90$ samples ($0.98$ and $0.95$). 
Overall, fewer than $100$ samples per benchmark suffice for reliable RSF extraction, a key insight for single-model attribution when only a limited number of target outputs are available. 
\begin{table}[!htbp]
\caption{Single-model attribution AUROC for different training set sizes.
Residual embeddings (extracted with a \SI{1}{\kilo\hertz} low-pass filter) are compared to target RSFs using the Mahalanobis distance.
Column headers indicate the number of samples used for RSF extraction.
}
\centering
\footnotesize
\begin{tabular}{lcccccc}
\toprule
Benchmarks & 70 & 80 & 90 & 100 & 110 & Full \\
\midrule
A-LJSpeech      & 0.98 & 1.00 & 1.00 & 1.00 & 1.00 & 1.00 \\
JSUT          & 0.96 & 0.98 & 0.99 & 1.00 & 1.00 & 1.00 \\
ASVspoof~LA   & 0.92 & 0.97 & 0.98 & 0.98 & 0.98 & 0.99 \\
CodecFake    & 0.92 & 0.95 & 0.95 & 0.96 & 0.97 & 0.98 \\
\bottomrule
\end{tabular}
\label{tab:sample_efficiency}
\end{table}

On an NVIDIA A40 GPU (\SI{48}{\giga\byte}), RSF extraction for A-LJSpeech takes only \SI{1.52}{\second} for $100$ samples and \SI{153.3}{\second} for the full $10{,}480$-sample set, while per-sample attribution requires just \SI{0.015}{\second}.
With computational complexity $O(N \cdot d^2)$ for extraction and $O(d^2)$ for attribution ($N$ training samples, $d$ RSF dimensionality), this demonstrates that RSF-based attribution is both efficient and scalable.

\subsection{Can Adaptive Attacks Evade RSF Attribution?}
While our primary threat model assumes a non-adaptive adversary, we evaluate three adaptive evasion attacks to stress-test attribution: one non-optimized baseline (RSF Swap) and two projected gradient descent (PGD) attacks~\cite{madry_2018ICLR}.
All are conducted in a closed-world setting, following the experimental setup from Sec.~\ref{sec:multi_model}, and assume the adversary to know all model–fingerprints and the defense mechanism.
The RSF Swap attack subtracts the source RSF and adds a random target RSF.
The first PGD variant iteratively minimizes the distance to the target RSF, while the second PGD variant adds an objective maximizing the distance from the source RSF, resulting in stronger attacks under limited perturbation budgets.
Each PGD attack runs for $1{,}000$ iterations per sample.
For evaluation, we use $100$ random synthetic test samples per category on ASVspoof~LA and CodecFake, each assigned a random target distinct from its ground truth.
An attack is deemed successful if the manipulated sample is attributed to the chosen target model.

Tab.~\ref{tab:evasion_attack} summarizes results.
The RSF Swap attack has limited success ($0.41$ for ASVspoof~LA and $0.70$ for CodecFake), showing that even full RSF knowledge does not guarantee misattribution.
The first PGD variant improves target attribution ($0.68$ for ASVspoof~LA and $0.90$ for CodecFake), while the second PGD typically achieves near-perfect success.
These results underscore the difficulty of defending against adaptive attacks, which remain challenging for many existing methods.
Notably, PGD attacks require full RSF access, and defenses like ensembles of multiple fingerprints per model could further hinder gradient-based optimization.
\begin{table}[!htbp]
\caption{
Success of adaptive evasion attacks.
Attribution assigns a sample to the model whose RSF has the smallest Mahalanobis distance.
}
\centering
\footnotesize
\begin{tabular}{lccc}
\toprule
Benchmarks & RSF Swap \quad & \quad PGD-1$^{*}$ & PGD-2$^{\dagger}$ \\
\midrule
ASVspoof~LA & 0.41 & \quad 0.68 & 0.97 \\
CodecFake    & 0.70 & \quad 0.90 & 1.00 \\
\bottomrule
\multicolumn{4}{l}{$^{*}$Minimizes the Mahalanobis distance ($s_{\operatorname{md}}$) to the target RSF.} \\
\multicolumn{4}{l}{$^{\dagger}$Extends PGD-1 by maximizing $s_{\operatorname{md}}$ from the source RSF.}
\end{tabular}
\label{tab:evasion_attack}
\end{table}

\subsection{Do Real-World Distortions Break RSF Attribution?}
We evaluate single-model attribution under realistic post-processing that may arise naturally or be applied adaptively by an attacker.
We consider MP3 compression, echo, reverberation, and background noise, and additionally test a lightweight defense that re-estimates RSFs on augmented training data.

\noindent \textbf{MP3 Compression:}
Inputs are encoded and decoded at $128\,\mathrm{kbit/s}$.
Compression reduces attribution performance, especially for JSUT and ASVspoof~LA (AUROC $0.57$–$0.64$; Tab.~\ref{tab:mp3-effect}).
Retraining on compressed data restores near-baseline performance across all benchmarks, showing that compression-induced shifts are largely recoverable through lightweight data augmentation.
\begin{table}[!htbp]
\caption{
Impact of MP3 compression on single-model attribution (AUROC).
Residual embeddings (\SI{1}{\kilo\hertz} low-pass) are compared to target RSFs using Mahalanobis distance.
}
\centering
\resizebox{\columnwidth}{!}{
\begin{tabular}{lcccc}
\toprule
MP3 compression & A-LJSpeech & JSUT  & ASVspoof~LA & CodecFake  \\
\midrule
No distortion  &  1.00 & 1.00 & 0.99 & 0.98 \\
$128\,\mathrm{kbit/s}$ & 0.63 & 0.57 & 0.64  & 0.79 \\
RSFs re-estimated$^{\dagger}$ &  0.99 & 1.00 & 0.97 & 0.98 \\
\bottomrule
\multicolumn{5}{l}{$^{\dagger}$ RSFs re-estimated on MP3-augmented data.}
\end{tabular}
}
\label{tab:mp3-effect}
\end{table}

\noindent \textbf{Echo distortion:}
Echo is simulated via a delay-and-add filter $\tilde{x}^{(i)}(k) = x^{(i)}(k) + \alpha x^{(i)}(k-D)$, where $D$ is the delay (corresponding to \SI{100}{\milli\second} or \SI{500}{\milli\second}) and $\alpha \in \{0.3, 0.5\}$ denotes the echo strength.
Mild echoes ($\alpha=0.3$) have limited impact, with AUROC remaining high across benchmarks and only small drops (down to $0.92$; Tab.~\ref{tab:echo-effect}).
Stronger echoes ($\alpha=0.5$, short delay) cause moderate degradation across multiple benchmarks, reaching $0.92$ for JSUT, ASVspoof~LA, and CodecFake.
Retraining on worst-case echo conditions restores baseline performance, indicating that even severe temporal distortions are largely recoverable through lightweight data augmentation.
\begin{table}[!htbp]
\caption{
Impact of echo distortion on single-model attribution (AUROC).
Residual embeddings (\SI{1}{\kilo\hertz} low-pass) are compared to target RSFs using the Mahalanobis distance.
}
\centering
\resizebox{\columnwidth}{!}{
\begin{tabular}{lcccc}
\toprule
Echo distortion & A-LJSpeech & JSUT  & ASVspoof~LA & CodecFake  \\
\midrule
No distortion  &  1.00 & 1.00 & 0.99 & 0.98 \\
$\alpha=0.3, D = \SI{100}{\milli\second}$ &  1.00 & 0.98 & 0.96 & 0.95 \\
$\alpha=0.3, D = \SI{500}{\milli\second}$ & 1.00 & 0.99 & 0.96 & 0.92 \\
$\alpha=0.5, D = \SI{100}{\milli\second}$ &  0.98 & 0.92 & 0.92 & 0.92 \\
$\alpha=0.5, D = \SI{500}{\milli\second}$ & 0.99 & 0.95 & 0.95 & 0.94 \\
RSFs re-estimated$^{\dagger}$ &  1.00 & 1.00 & 0.99 & 0.98 \\
\bottomrule
\multicolumn{5}{l}{$^{\dagger}$ RSFs re-estimated on worst-case echo augmented data ($\alpha=0.5$, $D=\SI{100}{\milli\second}$).}
\end{tabular}
}
\label{tab:echo-effect}
\end{table}

\noindent \textbf{Reverberation:}
We apply SpeechBrain's \texttt{AddReverb} using $26$ real room impulse responses from the $2014$ Reverberation Challenge~\cite{Ko_2017_ICASSP}.\footnote{AddReverb: \url{https://speechbrain-anonym.readthedocs.io/en/latest/API/speechbrain.processing.speech_augmentation.html}.}
Reverberation introduces the strongest degradation among realistic effects, reducing AUROC to $0.64$–$0.84$ across benchmarks (e.g., A-LJSpeech from $1.00$ to $0.65$; Tab.~\ref{tab:reverb-effect}).
However, retraining on reverberant data substantially improves performance (e.g., from $0.65$ to $0.97$), showing that acoustic mismatch is challenging yet largely recoverable through lightweight augmentation.
\begin{table}[!htbp]
\caption{
Impact of reverberation on single-model attribution (AUROC).
Residual embeddings (\SI{1}{\kilo\hertz} low-pass) are compared to target RSFs using the Mahalanobis distance.
}
\centering
\resizebox{\columnwidth}{!}{
\begin{tabular}{lcccc}
\toprule
Reverberation condition & A-LJSpeech & JSUT & ASVspoof~LA & CodecFake  \\
\midrule
No reverberation  &  1.00 & 1.00 & 0.99 & 0.98 \\
Reverberation & 0.65 & 0.64 & 0.81  & 0.84 \\
RSFs re-estimated$^{\dagger}$ &  0.97 & 0.96 & 0.91 & 0.92 \\
\bottomrule
\multicolumn{5}{l}{$^{\dagger}$ RSFs re-estimated on reverberation-augmented data.}
\end{tabular}
}
\label{tab:reverb-effect}
\end{table}

\noindent \textbf{Additive Background Noise:}
To evaluate robustness against additive noise, we use the MUSAN corpus~\cite{Ko_2017_ICASSP}, which includes music, speech, and environmental noise.
Noise is injected using SpeechBrain’s \texttt{EnvCorrupt},\footnote{EnvCorrupt: \url{https://speechbrain-anonym.readthedocs.io/en/latest/API/speechbrain.lobes.augment.html}.}
with parameters ranging from $10$ to $40$ (lower values correspond to stronger corruption).
Because these parameters are not directly interpretable, we report the corresponding average signal-to-noise ratio (SNR) in~\si{\decibel}~\cite{PIZZI_2026_CSLJournal}, where lower SNR indicates heavier noise.
Attribution remains highly robust even under severe corruption: AUROC is $0.89$ at approximately \SI{10}{\decibel} and already exceeds $0.94$ at \SI{18}{\decibel}, continuing to improve toward near-perfect performance at higher SNRs (Tab.~\ref{tab:background-noise}).
The smooth performance trend indicates that RSFs remain strongly discriminative under substantial additive noise.
\begin{table}[!htbp]
\caption{
Impact of additive background noise on single-model attribution (AUROC) on the A-LJSpeech benchmark.
Residual embeddings (\SI{1}{\kilo\hertz} low-pass) are compared to target RSFs using the Mahalanobis distance.
}
\centering
\resizebox{\columnwidth}{!}{
\begin{tabular}{cccccccc}
\toprule
SNR~(\si{\decibel}) & 10 & 14 & 18 & 23 & 27 & 32 & 37 \\
\midrule
AUROC & 0.89 & 0.92 & 0.94 & 0.96 & 0.97 & 0.98 & 0.99 \\
\bottomrule
\end{tabular}
}
\label{tab:background-noise}
\end{table}
\section{Conclusion}
With modern speech generation achieving high realism, detecting and attributing synthetic audio is critical for secure voice systems and forensic applications.
We introduced RSF, a lightweight and training-free framework that extracts model-specific artifacts from residuals extracted after filtering of speech signals.
By estimating the Mahalanobis distance of the residual of an audio sample to the residuals of a set of samples of a given model, RSF enables simple yet highly effective detection and attribution without requiring learned classifiers.
Across extensive experiments, RSF achieves near-perfect AUROC in open-world single-model attribution, reliably distinguishing samples generated by a target synthesis system from all other sources, including other synthetic models and real speech. 
In closed-world multi-model attribution, RSF provides strong discriminative cues, often matching or exceeding deep neural networks.
Moreover, it outperforms all baselines in out-of-domain detection, while avoiding the need for classifier retraining.
We further demonstrate robustness across languages, synthesis paradigms, and realistic audio distortions, where simple data augmentation restores performance under adverse conditions.
This combination of strong performance, minimal data requirements (fewer than $100$ samples), and training-free scalability makes RSF computationally efficient and aligned with Green AI principles.
Overall, our proposed framework provides a practical foundation for rapid-response detection, content moderation, and scalable forensic pipelines.
Future work will explore cross-lingual fingerprints and ensembles of multiple fingerprints, aiming to improve detection, attribution, and generalization.

\section{Acknowledgments}
Funded by the Deutsche Forschungsgemeinschaft (DFG,
German Research Foundation) under Germany’s Excellence Strategy- EXC 2092 CASA- 390781972 and by the Ministry of Culture and Science of North Rhine-Westphalia as part of the Lamarr Fellow Network.
We thank our colleagues Simon Heilig, Jonas Ricker, Jonas Thietke, and Tarini Saka, as well as our anonymous reviewers, for their valuable feedback.

\section{Generative AI Use Disclosure}
Generative AI tools were used solely for language editing, grammar correction, and improving the clarity of the manuscript.
%No generative AI tools were used to produce or design the scientific content, experiments, or results presented in this paper.
%All ideas, methods, and evaluations were developed by the authors, who take full responsibility for the content.

\bibliographystyle{IEEEtran}
\bibliography{camera_ready}

\clearpage

\begin{strip}
\begin{center}
    {\LARGE\bfseries Supplementary Material}\\[0.8em]

    {\large Lightweight Detection and Model Attribution of Synthetic Speech via Residual Statistical Fingerprints}\\[1em]

    Matías Pizarro$^1$,
    Mike Laszkiewicz$^2$,
    Dorothea Kolossa$^3$,
    Asja Fischer$^1$\\[0.8em]

    {\small
    $^1$ Faculty of Computer Science, Ruhr University Bochum, Germany\\
    $^2$ LKA NRW, Germany\\
    $^3$ Electronic Systems of Medical Engineering, Technische Universität Berlin, Germany
    }
\end{center}
\vspace{1em}
\end{strip}

% Your supplementary material starts here
This Supplementary Material contains additional experiments that support the findings presented in the main paper. 
These analyses were provided to the reviewers during the manuscript revision phase for final assessment and are not included in the official Interspeech proceedings.

\vspace{3ex}
\noindent{\large\bfseries A: Distance Metric and Spectral Filter Selection}
\vspace{1.5ex}
\noindent

\noindent (Section 4.2.) We compare correlation and Mahalanobis distance for open-world single-model attribution across multiple spectral filter configurations. 
Residual features are derived from STFT representations (\SI{8}{\milli\second} window, \SI{0.125}{\milli\second} hop) on the ASVspoof LA benchmark.
Two held-out systems (A16 and A19) are used to construct RSFs.

\noindent \textbf{Low-pass filter:}
We evaluate cutoff frequencies of $1$, $3$, and \SI{5}{\kilo\hertz}.
A \SI{1}{\kilo\hertz} cutoff achieves the highest average AUROC when paired with Mahalanobis distance ($0.96$ for A16, $1.00$ for A19), slightly outperforming higher cutoffs (Tab.~\ref{tab:lpf-conf}). 
We therefore adopt this configuration for low-pass filtering.
\begin{table}[!htbp]
\caption{
Single-model attribution AUROC on ASVspoof~LA using low-pass residuals. 
Rows denote source models and columns target models. 
Cells report correlation / Mahalanobis distance.
Bold marks the best average AUROC.
}
\centering
\resizebox{\columnwidth}{!}{
\begin{tabular}{lcccccc}
\toprule
 Source\textbackslash{}Target     & \multicolumn{2}{c}{Cutoff = \SI{1}{\kilo\hertz}} & \multicolumn{2}{c}{Cutoff = \SI{3}{\kilo\hertz}} & \multicolumn{2}{c}{Cutoff = \SI{5}{\kilo\hertz}} \\
        & A16 & A19$^{*}$ & A16 & A19$^{*}$ & A16 & A19$^{*}$ \\
\midrule
A07     & 0.67 / 0.82 & 0.97 / 0.99 & 0.67 / 0.81 & 0.97 / 0.99 & 0.63 / 0.78 & 0.93 / 0.99 \\
A08     & 0.97 / 1.00 & 1.00 / 1.00 & 0.97 / 1.00 & 1.00 / 1.00 & 0.95 / 1.00 & 1.00 / 1.00 \\
A09     & 0.99 / 1.00 & 1.00 / 1.00 & 0.99 / 1.00 & 1.00 / 1.00 & 0.99 / 1.00 & 1.00 / 1.00 \\
A10     & 0.94 / 1.00 & 1.00 / 1.00 & 0.96 / 1.00 & 1.00 / 1.00 & 0.96 / 1.00 & 1.00 / 1.00 \\
A11     & 0.93 / 1.00 & 1.00 / 1.00 & 0.95 / 1.00 & 1.00 / 1.00 & 0.96 / 1.00 & 1.00 / 1.00 \\
A12     & 0.95 / 1.00 & 1.00 / 1.00 & 0.97 / 1.00 & 1.00 / 1.00 & 0.98 / 1.00 & 1.00 / 1.00 \\
A13$^{\ddagger}$     & 0.99 / 1.00 & 1.00 / 1.00 & 0.99 / 1.00 & 1.00 / 1.00 & 0.99 / 1.00 & 1.00 / 1.00 \\
A14$^{\ddagger}$     & 0.98 / 1.00 & 1.00 / 1.00 & 0.99 / 1.00 & 1.00 / 1.00 & 0.99 / 1.00 & 1.00 / 1.00 \\
A15$^{\ddagger}$     & 0.97 / 1.00 & 1.00 / 1.00 & 0.99 / 1.00 & 1.00 / 1.00 & 0.98 / 1.00 & 1.00 / 1.00 \\
A16     & -      & 0.90 / 0.97 & -       & 0.91 / 0.98 & - & 0.87 / 0.98 \\
A17$^{*}$     & 0.89 / 1.00 & 1.00 / 1.00 & 0.93 / 1.00 & 1.00 / 1.00 & 0.96 / 1.00 & 1.00 / 1.00 \\
A18$^{*}$     & 0.48 / 0.99 & 0.86 / 0.98 & 0.57 / 0.99 & 0.89 / 0.97 & 0.71 / 0.99 & 0.91 / 0.99 \\
A19$^{*}$     & 0.18 / 0.66 & -       & 0.17 / 0.67 & -       & 0.25 / 0.67 & -       \\
Bonafide & 0.85 / 1.00 & 0.99 / 1.00 & 0.89 / 1.00 & 1.00 / 1.00 & 0.93 / 1.00 & 1.00 / 1.00 \\
\midrule
Avg.    & 0.84 / \textbf{0.96} & 0.98 / \textbf{1.00} & 0.86 / \textbf{0.95} & 0.98 / \textbf{0.99} & 0.87 / \textbf{0.95} & 0.97 / \textbf{0.99} \\
\bottomrule
\multicolumn{7}{l}{Unmarked: TTS only; $^{*}$ VC systems; $^{\ddagger}$ TTS+VC; higher AUROC indicates better separation.}
\end{tabular}
}
\label{tab:lpf-conf}
\end{table}

\noindent \textbf{High-pass filter:}
We evaluate cutoff frequencies of $1$, $3$, and \SI{5}{\kilo\hertz}.
Mahalanobis distance consistently outperforms correlation (Tab.~\ref{tab:hpf-conf}), but overall attribution performance is lower than with low-pass filtering.
\begin{table}[!htbp]
\caption{
Single-model attribution AUROC on ASVspoof~LA using high-pass residuals. 
Rows denote source models and columns target models. 
Cells report correlation / Mahalanobis distance.
Bold marks the best average AUROC.
}
\centering
\resizebox{\columnwidth}{!}{
\begin{tabular}{lcccccc}
\toprule
Source\textbackslash{}Target & \multicolumn{2}{c}{Cutoff = \SI{1}{\kilo\hertz}} & \multicolumn{2}{c}{Cutoff = \SI{3}{\kilo\hertz}} & \multicolumn{2}{c}{Cutoff = \SI{5}{\kilo\hertz}} \\
                & A16 & A19$^{*}$ & A16 & A19$^{*}$ & A16 & A19$^{*}$ \\
\midrule
A07      & 0.69 / 0.68 & 0.96 / 0.98 & 0.70 / 0.71 & 0.83 / 0.97 & 0.75 / 0.75 & 0.86 / 0.97 \\
A08      & 0.70 / 0.97 & 0.96 / 1.00 & 0.92 / 0.97 & 0.97 / 1.00 & 0.92 / 0.99 & 0.92 / 1.00 \\
A09      & 0.91 / 0.95 & 0.99 / 1.00 & 0.85 / 0.94 & 0.95 / 1.00 & 0.70 / 0.97 & 0.79 / 1.00 \\
A10      & 0.63 / 0.87 & 0.94 / 1.00 & 0.72 / 0.93 & 0.85 / 0.99 & 0.72 / 0.96 & 0.85 / 1.00 \\
A11      & 0.66 / 0.84 & 0.95 / 1.00 & 0.74 / 0.94 & 0.85 / 1.00 & 0.74 / 0.97 & 0.86 / 1.00 \\
A12      & 0.60 / 0.82 & 0.88 / 1.00 & 0.63 / 0.89 & 0.76 / 1.00 & 0.53 / 0.96 & 0.71 / 1.00 \\
A13$^{\ddagger}$      & 0.76 / 0.90 & 0.98 / 1.00 & 0.80 / 0.96 & 0.91 / 1.00 & 0.80 / 0.98 & 0.89 / 1.00 \\
A14$^{\ddagger}$      & 0.78 / 0.86 & 0.98 / 1.00 & 0.81 / 0.96 & 0.92 / 1.00 & 0.73 / 0.98 & 0.82 / 1.00 \\
A15$^{\ddagger}$      & 0.62 / 0.79 & 0.96 / 1.00 & 0.58 / 0.87 & 0.78 / 0.99 & 0.53 / 0.95 & 0.64 / 0.99 \\
A16      & -          & 0.79 / 0.98 & -          & 0.70 / 0.98 & -          & 0.64 / 0.97 \\
A17$^{*}$      & 0.35 / 0.80 & 0.70 / 0.96 & 0.39 / 0.85 & 0.46 / 0.94 & 0.41 / 0.94 & 0.46 / 0.97 \\
A18$^{*}$      & 0.53 / 0.59 & 0.47 / 0.88 & 0.41 / 0.68 & 0.48 / 0.89 & 0.36 / 0.78 & 0.50 / 0.95 \\
A19$^{*}$      & 0.60 / 0.55 & -          & 0.42 / 0.52 & -          & 0.43 / 0.60 & -          \\
Bonafide & 0.49 / 0.77 & 0.53 / 0.95 & 0.57 / 0.85 & 0.66 / 0.95 & 0.61 / 0.93 & 0.64 / 0.97 \\
\midrule
Avg. & 0.67 / \textbf{0.81} & 0.84 / \textbf{0.98} & 0.66 / \textbf{0.85} & 0.78 / \textbf{0.98} & 0.63 / \textbf{0.91} & 0.73 / \textbf{0.98} \\
\bottomrule
\multicolumn{7}{l}{Unmarked: TTS only; $^{*}$ VC systems; $^{\ddagger}$ TTS+VC; higher AUROC indicates better separation.}
\end{tabular}
}
\label{tab:hpf-conf}
\end{table}

\noindent \textbf{Band-stop filter:}
We evaluate stop bands of 4–7~\si{\kilo\hertz} and 5–6~\si{\kilo\hertz}.
The 4–7~\si{\kilo\hertz} stop band achieves the highest average AUROC with Mahalanobis distance ($0.90$ for A16, $0.98$ for A19; Tab.~\ref{tab:sbf-conf}), but still underperforms the low-pass configuration ($0.96$ and $1.00$).
\begin{table}[!htbp]
\caption{
Single-model attribution AUROC on ASVspoof~LA using band-stop residuals. 
Rows denote source models and columns target models. 
Cells report correlation / Mahalanobis distance.
Bold marks the best average AUROC.
}
\centering
\resizebox{\columnwidth}{!}{
\begin{tabular}{lcccc}
\toprule
Source\textbackslash{}Target & \multicolumn{2}{c}{Stop band = 4–7~\si{\kilo\hertz}} & \multicolumn{2}{c}{Stop band = 5–6~\si{\kilo\hertz}} \\
                & A16 & A19$^{*}$ & A16 & A19$^{*}$ \\
\midrule
A07      & 0.59 / 0.74 & 0.82 / 0.97 & 0.52 / 0.67 & 0.67 / 0.95 \\
A08      & 0.77 / 0.99 & 0.96 / 1.00 & 0.80 / 0.98 & 0.91 / 1.00 \\
A09      & 0.70 / 0.97 & 0.93 / 1.00 & 0.44 / 0.94 & 0.73 / 1.00 \\
A10      & 0.60 / 0.92 & 0.82 / 1.00 & 0.49 / 0.88 & 0.66 / 0.99 \\
A11      & 0.61 / 0.96 & 0.84 / 1.00 & 0.61 / 0.91 & 0.77 / 1.00 \\
A12      & 0.51 / 0.95 & 0.82 / 1.00 & 0.46 / 0.93 & 0.75 / 1.00 \\
A13$^{\ddagger}$      & 0.68 / 0.97 & 0.88 / 1.00 & 0.57 / 0.95 & 0.71 / 1.00 \\
A14$^{\ddagger}$      & 0.61 / 0.96 & 0.88 / 1.00 & 0.43 / 0.92 & 0.70 / 1.00 \\
A15$^{\ddagger}$      & 0.50 / 0.95 & 0.81 / 1.00 & 0.46 / 0.91 & 0.64 / 1.00 \\
A16      & -         & 0.80 / 0.97 & -          & 0.72 / 0.96 \\
A17$^{*}$      & 0.37 / 0.96 & 0.54 / 0.99 & 0.51 / 0.92 & 0.54 / 0.99 \\
A18$^{*}$      & 0.49 / 0.79 & 0.48 / 0.83 & 0.38 / 0.61 & 0.36 / 0.71 \\
A19$^{*}$      & 0.38 / 0.59 & -          & 0.57 / 0.51 & -          \\
Bonafide & 0.38 / 0.95 & 0.61 / 0.99 & 0.51 / 0.89 & 0.54 / 0.98 \\
\midrule
Avg. & 0.55 / \textbf{0.90} & 0.79 / \textbf{0.98} & 0.50 / \textbf{0.85} & 0.68 / \textbf{0.97} \\
\bottomrule
\multicolumn{5}{l}{Unmarked: TTS only; $^{*}$ VC systems; $^{\ddagger}$ TTS+VC.}
\end{tabular}
}
\label{tab:sbf-conf}
\end{table}

\noindent \textbf{Band-pass filter:}
We evaluate pass bands of 4–7~\si{\kilo\hertz} and 5–6~\si{\kilo\hertz}.
Band-pass filtering achieves performance comparable to low-pass filtering (Tab.~\ref{tab:bpf}), with similarly high AUROC across both targets.
\begin{table}[!htbp]
\caption{
Single-model attribution AUROC on ASVspoof~LA using band-pass residuals. 
Rows denote source models and columns target models. 
Cells report correlation / Mahalanobis distance.
Bold marks the best average AUROC.
}
\centering
\resizebox{\columnwidth}{!}{
\begin{tabular}{lcccc}
\toprule
Source\textbackslash{}Target & \multicolumn{2}{c}{Pass band = 4–7~\si{\kilo\hertz}} & \multicolumn{2}{c}{Pass band = 5–6~\si{\kilo\hertz}} \\
                & A16 & A19$^{*}$ & A16 & A19$^{*}$ \\
\midrule
A07      & 0.64 / 0.78 & 0.89 / 0.98 & 0.62 / 0.76 & 0.89 / 0.98 \\
A08      & 0.90 / 1.00 & 0.98 / 1.00 & 0.91 / 1.00 & 0.99 / 1.00 \\
A09      & 0.98 / 1.00 & 1.00 / 1.00 & 0.97 / 1.00 & 1.00 / 1.00 \\
A10      & 0.97 / 1.00 & 0.99 / 1.00 & 0.95 / 1.00 & 0.99 / 1.00 \\
A11      & 0.97 / 1.00 & 0.99 / 1.00 & 0.94 / 1.00 & 0.98 / 1.00 \\
A12      & 0.99 / 1.00 & 1.00 / 1.00 & 0.97 / 1.00 & 0.99 / 1.00 \\
A13$^{\ddagger}$      & 1.00 / 1.00 & 1.00 / 1.00 & 0.99 / 1.00 & 1.00 / 1.00 \\
A14$^{\ddagger}$      & 1.00 / 1.00 & 1.00 / 1.00 & 0.99 / 1.00 & 1.00 / 1.00 \\
A15$^{\ddagger}$      & 0.99 / 1.00 & 0.99 / 1.00 & 0.98 / 1.00 & 1.00 / 1.00 \\
A16      & -          & 0.84 / 0.98 & -          & 0.83 / 0.98 \\
A17$^{*}$      & 0.98 / 1.00 & 0.99 / 1.00 & 0.95 / 1.00 & 0.98 / 1.00 \\
A18$^{*}$      & 0.88 / 1.00 & 0.92 / 0.99 & 0.80 / 1.00 & 0.86 / 0.98 \\
A19$^{*}$      & 0.34 / 0.64 & -          & 0.35 / 0.64 & -          \\
Bonafide & 0.96 / 1.00 & 0.99 / 1.00 & 0.92 / 1.00 & 0.98 / 1.00 \\
\midrule
Avg. & 0.88 / \textbf{0.95} & 0.97 / \textbf{1.00} & 0.88 / \textbf{0.97} & 0.97 / \textbf{0.99} \\
\bottomrule
\multicolumn{5}{l}{Unmarked: TTS only; $^{*}$ VC systems; $^{\ddagger}$ TTS+VC.}
\end{tabular}
}
\label{tab:bpf}
\end{table}

We therefore adopt a \SI{1}{\kilo\hertz} low-pass filter and a 5–6~\si{\kilo\hertz} band-pass filter, both yielding strong attribution performance.

\newpage

\vspace{3ex}
\noindent{\large\bfseries B: STFT Resolution Selection}
\vspace{1.5ex}
\noindent

\noindent (Section 4.3.) Pairwise AUROCs for the \SI{25}{\milli\second} / \SI{10}{\milli\second} STFT configuration are reported for A-LJSpeech (Tab.~\ref{tab:pairwise-auroc}), JSUT (Tab.~\ref{tab:merged_auroc_mbmg_pwg_full}), ASVspoof~LA (Tab.~\ref{tab:pairwise-a01-a06}), and CodecFake (Tab.~\ref{tab:pairwise-c1-c6}). 
For the selected \SI{8}{\milli\second} / \SI{0.125}{\milli\second} STFT configuration, full matrices are provided for A-LJSpeech (Tab.~\ref{tab:merged_auroc_full_all}) and JSUT (Tab.~\ref{tab:merged_auroc_mbmg_pwg_full_2}), while results for ASVspoof~LA and CodecFake appear in the main text (Section 5.1).
\\
\begin{table*}[t]
\caption{
Pairwise single-model attribution AUROC on A-LJSpeech using Mahalanobis distance with the \SI{25}{\milli\second} / \SI{10}{\milli\second} STFT configuration. 
Rows correspond to source models and columns to target models. 
Results are reported for \SI{1}{\kilo\hertz} low-pass and 5–6~\si{\kilo\hertz} band-pass filtering.$^{\dagger}$
}
\centering
\resizebox{\textwidth}{!}{
\begin{tabular}{lcccccccccc}
\toprule
Source\textbackslash{}Target & FastDiff & ProDiff & MG-L  & Avo   & BVG   & HF-G  & MB-MG & PWG   & WGlow & NSF   \\
\midrule
FastDiff   & -        & 0.96 / 0.96 & 1.00 / 1.00 & 1.00 / 1.00 & 1.00 / 1.00 & 1.00 / 1.00 & 1.00 / 1.00 & 1.00 / 1.00 & 1.00 / 1.00 & 1.00 / 1.00 \\
ProDiff    & 0.98 / 0.99 & -        & 1.00 / 1.00 & 1.00 / 1.00 & 1.00 / 1.00 & 1.00 / 1.00 & 1.00 / 1.00 & 0.99 / 1.00 & 1.00 / 1.00 & 1.00 / 1.00 \\
MG-L       & 1.00 / 1.00 & 1.00 / 1.00 & -        & 1.00 / 1.00 & 1.00 / 1.00 & 1.00 / 1.00 & 1.00 / 1.00 & 1.00 / 1.00 & 1.00 / 1.00 & 1.00 / 1.00 \\
Avo        & 1.00 / 1.00 & 1.00 / 1.00 & 1.00 / 1.00 & -        & 1.00 / 0.99 & 1.00 / 1.00 & 0.98 / 0.97 & 0.96 / 0.93 & 0.99 / 0.98 & 0.97 / 0.96 \\
BVG        & 1.00 / 1.00 & 1.00 / 1.00 & 1.00 / 1.00 & 1.00 / 0.99 & -        & 1.00 / 1.00 & 1.00 / 1.00 & 1.00 / 1.00 & 1.00 / 1.00 & 1.00 / 1.00 \\
HF-G       & 1.00 / 1.00 & 1.00 / 1.00 & 1.00 / 1.00 & 1.00 / 1.00 & 1.00 / 1.00 & -        & 1.00 / 1.00 & 1.00 / 1.00 & 1.00 / 0.99 & 1.00 / 1.00 \\
MB-MG      & 1.00 / 1.00 & 1.00 / 1.00 & 1.00 / 1.00 & 0.97 / 0.96 & 1.00 / 1.00 & 1.00 / 1.00 & -        & 0.97 / 0.96 & 1.00 / 1.00 & 1.00 / 1.00 \\
PWG        & 1.00 / 1.00 & 1.00 / 1.00 & 1.00 / 1.00 & 0.95 / 0.92 & 1.00 / 0.99 & 1.00 / 1.00 & 0.98 / 0.98 & -        & 0.99 / 0.98 & 0.94 / 0.93 \\
WGlow      & 1.00 / 1.00 & 1.00 / 1.00 & 1.00 / 1.00 & 1.00 / 1.00 & 1.00 / 1.00 & 1.00 / 0.99 & 1.00 / 1.00 & 1.00 / 1.00 & -        & 1.00 / 1.00 \\
NSF        & 1.00 / 1.00 & 1.00 / 1.00 & 1.00 / 1.00 & 0.98 / 0.97 & 1.00 / 1.00 & 1.00 / 1.00 & 1.00 / 1.00 & 0.95 / 0.93 & 1.00 / 0.99 & -        \\
Real       & 1.00 / 1.00 & 1.00 / 1.00 & 1.00 / 1.00 & 0.98 / 0.97 & 1.00 / 1.00 & 1.00 / 1.00 & 0.97 / 0.96 & 0.98 / 0.97 & 1.00 / 1.00 & 1.00 / 1.00 \\
\midrule
Avg.       & 1.00 / 1.00 & 1.00 / 1.00 & 1.00 / 1.00 & 0.99 / 0.98 & 1.00 / 1.00 & 1.00 / 1.00 & 0.99 / 0.99 & 0.98 / 0.98 & 1.00 / 0.99 & 0.99 / 0.99 \\
\bottomrule
\multicolumn{11}{l}{$^{\dagger}$ Higher AUROC indicates better separation.}
\end{tabular}
}
\label{tab:pairwise-auroc}
\end{table*}
\begin{table*}[t]
\caption{
Pairwise single-model attribution AUROC on A-LJSpeech using Mahalanobis distance with the \SI{8}{\milli\second} / \SI{0.125}{\milli\second} STFT configuration. 
Rows correspond to source models and columns to target models. 
Results are reported for \SI{1}{\kilo\hertz} low-pass and 5–6~\si{\kilo\hertz} band-pass filtering.$^{\dagger}$
}
\centering
\resizebox{\textwidth}{!}{
\begin{tabular}{lcccccccccc}
\toprule
Source\textbackslash{}Target & FastDiff & ProDiff & MG-L & Avo & BVG & HF-G & MB-MG & PWG & WGlow & NSF \\
\midrule
FastDiff & - & 1.00 / 1.00 & 1.00 / 1.00 & 1.00 / 1.00 & 1.00 / 1.00 & 1.00 / 1.00 & 1.00 / 1.00 & 1.00 / 1.00 & 1.00 / 1.00 & 1.00 / 1.00 \\
ProDiff  & 1.00 / 1.00 & - & 1.00 / 1.00 & 1.00 / 1.00 & 1.00 / 1.00 & 1.00 / 1.00 & 1.00 / 1.00 & 1.00 / 1.00 & 1.00 / 1.00 & 1.00 / 1.00 \\
MG-L     & 1.00 / 1.00 & 1.00 / 1.00 & - & 1.00 / 1.00 & 1.00 / 1.00 & 1.00 / 1.00 & 1.00 / 1.00 & 1.00 / 1.00 & 1.00 / 1.00 & 1.00 / 1.00 \\
Avo      & 1.00 / 1.00 & 1.00 / 1.00 & 1.00 / 1.00 & - & 1.00 / 1.00 & 1.00 / 1.00 & 1.00 / 1.00 & 1.00 / 1.00 & 1.00 / 1.00 & 1.00 / 1.00 \\
BVG      & 1.00 / 1.00 & 1.00 / 1.00 & 1.00 / 1.00 & 1.00 / 1.00 & - & 1.00 / 1.00 & 1.00 / 1.00 & 1.00 / 1.00 & 1.00 / 1.00 & 1.00 / 1.00 \\
HF-G     & 1.00 / 1.00 & 1.00 / 1.00 & 1.00 / 1.00 & 1.00 / 1.00 & 1.00 / 1.00 & - & 1.00 / 1.00 & 1.00 / 1.00 & 1.00 / 1.00 & 1.00 / 1.00 \\
MB-MG    & 1.00 / 1.00 & 1.00 / 1.00 & 1.00 / 1.00 & 1.00 / 1.00 & 1.00 / 1.00 & 1.00 / 1.00 & - & 1.00 / 1.00 & 1.00 / 1.00 & 1.00 / 1.00 \\
PWG      & 1.00 / 1.00 & 1.00 / 1.00 & 1.00 / 1.00 & 0.99 / 0.99 & 0.99 / 0.99 & 1.00 / 1.00 & 1.00 / 1.00 & - & 1.00 / 1.00 & 0.99 / 0.99 \\
WGlow    & 1.00 / 1.00 & 1.00 / 1.00 & 1.00 / 1.00 & 1.00 / 1.00 & 1.00 / 1.00 & 1.00 / 1.00 & 1.00 / 1.00 & 1.00 / 1.00 & - & 1.00 / 1.00 \\
NSF      & 1.00 / 1.00 & 1.00 / 1.00 & 1.00 / 1.00 & 1.00 / 1.00 & 1.00 / 1.00 & 1.00 / 1.00 & 1.00 / 1.00 & 1.00 / 0.99 & 1.00 / 1.00 & - \\
Real     & 1.00 / 1.00 & 1.00 / 1.00 & 1.00 / 1.00 & 1.00 / 1.00 & 1.00 / 1.00 & 1.00 / 1.00 & 1.00 / 1.00 & 1.00 / 1.00 & 1.00 / 1.00 & 1.00 / 1.00 \\
\midrule
Avg.     & 1.00 / 1.00 & 1.00 / 1.00 & 1.00 / 1.00 & 1.00 / 1.00 & 1.00 / 1.00 & 1.00 / 1.00 & 1.00 / 1.00 & 1.00 / 1.00 & 1.00 / 1.00 & 1.00 / 1.00 \\
\bottomrule
\multicolumn{11}{l}{$^{\dagger}$ Higher AUROC indicates better separation.}
\end{tabular}
}
\label{tab:merged_auroc_full_all}
\end{table*}
\begin{table}[!tbp]
\caption{
Pairwise single-model attribution AUROC on JSUT using Mahalanobis distance with the \SI{25}{\milli\second} / \SI{10}{\milli\second} STFT configuration. 
Rows correspond to source models and columns to target models. 
Results are reported for \SI{1}{\kilo\hertz} low-pass and 5–6~\si{\kilo\hertz} band-pass filtering.$^{\dagger}$
}
\centering
\begin{tabular}{lcc}
\toprule
Source\textbackslash{}Target & MB-MG & PWG \\
\midrule
MB-MG & - & 1.00 / 1.00 \\
PWG   & 1.00 / 1.00 & - \\
Real  & 1.00 / 1.00 & 1.00 / 1.00 \\
\midrule
Avg.  & 1.00 / 1.00 & 1.00 / 1.00 \\
\bottomrule
\multicolumn{3}{l}{$^{\dagger}$ Higher AUROC indicates better separation.}
\end{tabular}
\label{tab:merged_auroc_mbmg_pwg_full}
\end{table}
\begin{table}[!htbp]
\caption{
Pairwise single-model attribution AUROC on JSUT using Mahalanobis distance with the \SI{8}{\milli\second} / \SI{0.125}{\milli\second} STFT configuration. 
Rows correspond to source models and columns to target models. 
Results are reported for \SI{1}{\kilo\hertz} low-pass and 5–6~\si{\kilo\hertz} band-pass filtering.$^{\dagger}$
}
\centering
\begin{tabular}{lcc}
\toprule
Source\textbackslash{}Target & MB-MG & PWG \\
\midrule
MB-MG & - & 1.00 / 1.00 \\
PWG   & 1.00 / 1.00 & - \\
Real  & 1.00 / 1.00 & 1.00 / 1.00 \\
\midrule
Avg.  & 1.00 / 1.00 & 1.00 / 1.00 \\
\bottomrule
\multicolumn{3}{l}{$^{\dagger}$ Higher AUROC indicates better separation.}
\end{tabular}
\label{tab:merged_auroc_mbmg_pwg_full_2}
\end{table}
\begin{table}[!htbp]
\caption{
Pairwise single-model attribution AUROC on ASVspoof~LA using Mahalanobis distance with the \SI{25}{\milli\second} / \SI{10}{\milli\second} STFT configuration. 
Rows correspond to source models and columns to target models. 
Results are reported for \SI{1}{\kilo\hertz} low-pass and 5–6~\si{\kilo\hertz} band-pass filtering.$^{\dagger}$
}
\centering
\resizebox{\columnwidth}{!}{
\begin{tabular}{lcccccc}
\toprule
Source\textbackslash{}Target & A01           & A02           & A03           & A04           & A05$^{*}$           & A06$^{*}$          \\
\midrule
A01       & -             & 1.00 / 1.00   & 1.00 / 1.00   & 0.80 / 0.76   & 1.00 / 1.00   & 1.00 / 1.00   \\
A02       & 0.94 / 0.98   & -             & 0.44 / 0.87   & 0.88 / 0.98   & 0.77 / 0.74   & 1.00 / 1.00   \\
A03       & 0.98 / 1.00   & 0.99 / 1.00   & -             & 0.96 / 0.98   & 0.91 / 0.95   & 1.00 / 1.00   \\
A04       & 0.84 / 0.92   & 1.00 / 1.00   & 1.00 / 1.00   & -             & 1.00 / 1.00   & 0.99 / 1.00   \\
A05$^{*}$       & 0.97 / 0.98   & 0.99 / 0.99   & 0.72 / 0.95   & 0.88 / 0.98   & -             & 1.00 / 1.00   \\
A06$^{*}$       & 0.84 / 0.82   & 1.00 / 1.00   & 1.00 / 1.00   & 0.71 / 0.71   & 1.00 / 1.00   & -             \\
A07       & 0.85 / 0.79   & 1.00 / 1.00   & 1.00 / 1.00   & 0.81 / 0.77   & 1.00 / 1.00   & 0.99 / 0.99   \\
A08       & 1.00 / 1.00   & 1.00 / 1.00   & 1.00 / 1.00   & 1.00 / 1.00   & 1.00 / 1.00   & 1.00 / 1.00   \\
A09       & 0.99 / 1.00   & 1.00 / 1.00   & 0.81 / 0.97   & 0.98 / 0.98   & 0.90 / 0.99   & 1.00 / 1.00   \\
A10       & 0.99 / 0.99   & 1.00 / 1.00   & 0.90 / 0.92   & 0.98 / 0.98   & 0.95 / 0.95   & 1.00 / 1.00   \\
A11       & 1.00 / 1.00   & 1.00 / 1.00   & 0.98 / 0.99   & 0.99 / 0.99   & 0.99 / 0.99   & 1.00 / 1.00   \\
A12       & 1.00 / 1.00   & 1.00 / 1.00   & 0.92 / 0.98   & 0.97 / 0.98   & 0.95 / 1.00   & 1.00 / 1.00   \\
A13$^{\ddagger}$       & 1.00 / 1.00   & 1.00 / 1.00   & 0.92 / 0.94   & 0.99 / 0.99   & 0.96 / 0.97   & 1.00 / 1.00   \\
A14$^{\ddagger}$       & 0.99 / 0.99   & 1.00 / 1.00   & 0.86 / 0.90   & 0.99 / 0.99   & 0.94 / 0.95   & 1.00 / 1.00   \\
A15$^{\ddagger}$       & 1.00 / 1.00   & 1.00 / 1.00   & 0.89 / 0.91   & 0.98 / 0.98   & 0.90 / 0.91   & 1.00 / 1.00   \\
A17$^{*}$       & 0.99 / 0.98   & 1.00 / 1.00   & 0.92 / 0.97   & 0.95 / 0.97   & 0.88 / 0.91   & 1.00 / 1.00   \\
A18$^{*}$       & 0.95 / 0.92   & 1.00 / 1.00   & 0.83 / 0.90   & 0.84 / 0.87   & 0.91 / 0.94   & 0.95 / 0.98   \\
Bonafide  & 0.99 / 0.99   & 1.00 / 1.00   & 0.95 / 0.98   & 0.95 / 0.96   & 0.95 / 0.95   & 1.00 / 1.00   \\
\midrule
Avg.      & 0.96 / 0.96   & 1.00 / 1.00   & 0.89 / 0.96   & 0.92 / 0.93   & 0.94 / 0.96   & 1.00 / 1.00   \\
\bottomrule
\multicolumn{7}{l}{Unmarked: TTS only; $^{*}$ VC systems; $^{\ddagger}$ TTS+VC; higher AUROC indicates better separation.}

\end{tabular}
}
\label{tab:pairwise-a01-a06}
\end{table}
\begin{table}[!htbp]
\caption{
Pairwise single-model attribution AUROC on CodecFake using Mahalanobis distance with the \SI{25}{\milli\second} / \SI{10}{\milli\second} STFT configuration. 
Rows correspond to source models and columns to target models. 
Results are reported for \SI{1}{\kilo\hertz} low-pass and 5–6~\si{\kilo\hertz} band-pass filtering.$^{\dagger}$
}
\centering
\resizebox{\columnwidth}{!}{
\begin{tabular}{lcccccc}
\toprule
Source\textbackslash{}Target & C1           & C2           & C3           & C4           & C5           & C6           \\
\midrule
C1     & -            & 1.00 / 1.00  & 1.00 / 1.00  & 1.00 / 1.00  & 1.00 / 1.00  & 1.00 / 1.00  \\
C2     & 0.99 / 0.99  & -            & 1.00 / 1.00  & 0.98 / 0.99  & 0.99 / 0.99  & 0.88 / 0.92  \\
C3     & 1.00 / 1.00  & 1.00 / 1.00  & -            & 1.00 / 1.00  & 1.00 / 1.00  & 1.00 / 1.00  \\
C4     & 1.00 / 1.00  & 0.99 / 0.99  & 1.00 / 1.00  & -            & 1.00 / 1.00  & 0.99 / 1.00  \\
C5     & 1.00 / 1.00  & 0.96 / 0.96  & 1.00 / 1.00  & 0.99 / 0.99  & -            & 0.95 / 0.97  \\
C6     & 0.99 / 0.98  & 0.84 / 0.83  & 1.00 / 1.00  & 0.97 / 0.98  & 0.97 / 0.97  & -            \\
C7     & 1.00 / 0.99  & 0.92 / 0.93  & 1.00 / 1.00  & 0.98 / 0.99  & 0.99 / 0.99  & 0.89 / 0.92  \\
Real   & 1.00 / 1.00  & 0.97 / 0.97  & 1.00 / 1.00  & 1.00 / 1.00  & 1.00 / 1.00  & 0.97 / 0.97  \\
\midrule
Avg.   & 1.00 / 0.99  & 0.96 / 0.96  & 1.00 / 1.00  & 0.99 / 0.99  & 0.99 / 0.99  & 0.95 / 0.97  \\
\bottomrule
\multicolumn{7}{l}{$^{\dagger}$ Higher AUROC indicates better separation.}
\end{tabular}
}
\label{tab:pairwise-c1-c6}
\end{table}

\vspace{3ex}
\noindent{\large\bfseries C: Closed-World Multi-Model Attribution}
\vspace{1.5ex}
\noindent

\noindent (Section 5.2.) Tab.~\ref{tab:closed-world-prec} reports Precision and Recall for the closed-world multi-model attribution task, and Tab.~\ref{tab:closed-world-std} presents the corresponding standard deviations.
All values are averaged over five independent runs with ten training epochs per run.
\\

\begin{table}[!htbp]
\caption{
Closed-world multi-model attribution performance across benchmarks.
Cells: Precision / Recall, averaged over five runs with ten training epochs each.$^{\dagger}$
}
\centering
\resizebox{\columnwidth}{!}{%
\begin{tabular}{lcccc}
\toprule
Model & A-LJSpeech & JSUT & ASVspoof~LA & CodecFake \\
\midrule
X-vector        & 0.99 / 0.99 & 0.99 / 0.99 & \textbf{1.00 / 1.00} & \textbf{1.00 / 1.00} \\
LCNN            & 0.98 / 0.98 & 0.98 / 0.98 & \textbf{1.00 / 1.00} & 0.99 / 0.98 \\
ResNet          & 0.98 / 0.98 & 0.99 / 0.99 & \textbf{1.00 / 1.00} & \textbf{1.00 / 1.00} \\
SE-ResNet       & 0.98 / 0.98 & \textbf{1.00 / 1.00} & \textbf{1.00 / 1.00} & \textbf{1.00 / 1.00} \\
RSF (ours)      & \textbf{1.00 / 1.00} & \textbf{1.00 / 1.00} & 0.97 / 0.97 & 0.99 / 0.99 \\
RSF CNN (ours)  & \textbf{1.00 / 1.00} & \textbf{1.00 / 1.00} & \textbf{1.00 / 1.00} & \textbf{1.00 / 1.00} \\
\bottomrule
\multicolumn{5}{l}{$^{\dagger}$ Bold marks best-performing models per column.}
\end{tabular}
}
\label{tab:closed-world-prec}
\end{table}
\begin{table}[!htbp]
\caption{
Closed-world multi-model attribution performance variability across benchmarks.
Cells: standard deviation of Accuracy / F1 score, averaged over five runs with ten training epochs.$^{\dagger}$
}
\centering
\resizebox{\columnwidth}{!}{%
\begin{tabular}{lcccc}
\toprule
Model & A-LJSpeech & JSUT & ASVspoof~LA & CodecFake \\
\midrule
X-vector        & 0.0027 / 0.0027 & 0.0033 / 0.0033 & \textbf{0.0003 / 0.0003} & 0.0010 / 0.0010 \\
LCNN            & 0.0092 / 0.0092 & 0.0157 / 0.0157 & 0.0011 / 0.0011 & 0.0028 / 0.0028 \\
ResNet          & 0.0080 / 0.0081 & 0.0150 / 0.0150 & 0.0005 / 0.0005 & 0.0031 / 0.0031 \\
SE-ResNet       & 0.0109 / 0.0113 & 0.0035 / 0.0035 & \textbf{0.0003 / 0.0003} & 0.0016 / 0.0016 \\
RSF (ours)      & \textbf{0.0001 / 0.0001} & \textbf{0.0000 / 0.0000} & 0.0010 / 0.0010 & 0.0010 / 0.0010 \\
RSF CNN (ours)  & 0.0002 / 0.0002 & 0.0081 / 0.0081 & 0.0014 / 0.0014 & \textbf{0.0003 / 0.0003} \\
\bottomrule
\multicolumn{5}{l}{$^{\dagger}$ Bold marks lowest (best) variability per column.}
\end{tabular}
}
\label{tab:closed-world-std}
\end{table}

\vspace{3ex}
\noindent{\large\bfseries D: Real vs. Synthetic Classification}
\vspace{1.5ex}
\noindent

\noindent (Section 5.3.) Tab.~\ref{tab:fake-vs-real-prec} reports Precision and Recall for the real vs.~synthetic classification task, and Tab.~\ref{tab:fake-vs-real-std} presents the corresponding standard deviations.
All values are averaged over five independent runs with ten training epochs per run.
\begin{table}[!htbp]
\caption{
Real vs.~Synthetic classification performance across benchmarks.
Cells: Precision / Recall, averaged over five runs with ten training epochs each.$^{\dagger}$
}
\centering
\resizebox{\columnwidth}{!}{%
\begin{tabular}{lcccc}
\toprule
Model & A-LJSpeech & JSUT & ASVspoof & CodecFake \\
\midrule
X-vector        & 0.88 / 0.99 & \textbf{1.00 / 1.00} & 0.95 / 0.80 & \textbf{1.00} / 0.86 \\
LCNN            & 0.96 / 0.97 & 0.98 / 1.00 & 0.97 / 0.89 & 0.95 / \textbf{0.92} \\
ResNet          & 0.98 / 0.99 & \textbf{1.00 / 1.00} & 0.99 / 0.90 & \textbf{1.00} / 0.87 \\
SE-ResNet       & 0.98 / 0.99 & \textbf{1.00 / 1.00} & 0.99 / 0.90 & \textbf{1.00} / 0.87 \\
RSF CNN (ours)  & \textbf{1.00 / 1.00} & \textbf{1.00 / 1.00} & \textbf{1.00 / 0.91} & \textbf{1.00} / 0.87 \\
\bottomrule
\multicolumn{5}{l}{$^{\dagger}$ Bold marks best-performing models per column.}
\end{tabular}
}
\label{tab:fake-vs-real-prec}
\end{table}
\begin{table}[!htbp]
\caption{
Real vs.~Synthetic classification performance variability across benchmarks.
Cells: standard deviation of Accuracy / F1 score, averaged over five runs with ten training epochs.
Bold marks lowest variability per column.
}
\centering
\resizebox{\columnwidth}{!}{%
\begin{tabular}{lcccc}
\toprule
Model & A-LJSpeech & JSUT & ASVspoof & CodecFake \\
\midrule
X-vector        & 0.1367 / 0.0991 & \textbf{0.0000 / 0.0000} & 0.0333 / 0.0470 & \textbf{0.0022 / 0.0025} \\
LCNN            & 0.0077 / 0.0073 & 0.0228 / 0.0216 & 0.0140 / 0.0156 & 0.0197 / 0.0179 \\
ResNet          & \textbf{0.0013 / 0.0012} & 0.0009 / 0.0009 & 0.0076 / 0.0086 & 0.0059 / 0.0067 \\
SE-ResNet       & 0.0046 / 0.0044 & \textbf{0.0000 / 0.0000} & \textbf{0.0073 / 0.0081} & 0.0070 / 0.0081 \\
RSF CNN (ours)  & 0.0028 / 0.0028 & 0.0004 / 0.0004 & 0.0143 / 0.0158 & 0.0029 / 0.0033 \\
\bottomrule
\end{tabular}
}
\label{tab:fake-vs-real-std}
\end{table}

\end{document}